\documentclass[a4paper,fleqn,usenatbib]{mnras}

\usepackage{newtxtext,newtxmath}

\usepackage[T1]{fontenc}

\DeclareRobustCommand{\VAN}[3]{#2}
\let\VANthebibliography\thebibliography
\def\thebibliography{\DeclareRobustCommand{\VAN}[3]{##3}\VANthebibliography}

\usepackage{graphicx}	
\usepackage{amsmath}	

\usepackage[dvipsnames]{xcolor}

\newcommand{\vecF}{\mbox{\boldmath $F$} {}}
\newcommand{\vecr}{\mbox{\boldmath $r$} {}}

\newcommand{\vecJ}{\mbox{\boldmath $J$} {}}
\newcommand{\vecV}{\mbox{\boldmath $V$} {}}

\title[Gaps by retrograde satellites]{Structure of gaps induced by retrograde satellites embedded in accretion discs}

\author[F. J. S\'anchez-Salcedo  and A. Santill\'an]{F. J. S\'anchez-Salcedo$^{1}$\thanks{E-mail: jsanchez@astro.unam.mx} and A. Santill\'an$^{2}$\\
$^{1}$Universidad Nacional Aut\'onoma
de M\'exico, Instituto de Astronom\'{\i}a,
A.P. 70-264,  04510, Ciudad de M\'exico, Mexico\\
$^{2}$Departamento de Superc\'omputo, Direcci\'on General de C\'omputo y de Tecnolog\'{\i}as de Informaci\'on y
Comunicaci\'on, Universidad Nacional\\ Aut\'onoma de M\'exico, Ciudad Universitaria, 04510, Ciudad de M\'exico, Mexico
}

\date{Accepted XXX. Received YYY; in original form ZZZ}

\pubyear{2024}

\begin{document}
\label{firstpage}
\pagerange{\pageref{firstpage}--\pageref{lastpage}}
\maketitle

\begin{abstract}
Using 2D simulations, we investigate how a non-accreting satellite
on a fixed retrograde circular orbit affects the structure of the accretion disc in which 
it is embedded. We vary the satellite-to-primary mass ratio $q$, 
the disc viscosity $\nu$, and the inner boundary conditions.
A viscous criterion for gap opening is derived, which is broadly consistent 
with the simulations. We find a scaling relation of the gap depth with
$q$ and $\nu$. Unlike the prograde case, the satellite is located 
at the gap's inner edge, resulting in a surface density at the satellite's orbital radius up 
to $20$ times higher than at the gap's minimum. As the viscosity decreases,
the gap depth increases, while the radial shift of the gap and the satellite's orbital 
radius decreases. Gap-opening satellites may drive radial  motions in the disc, producing 
eccentric gaps.
Positioned at the gap edge, satellites experience a rapidly fluctuating environment.
Migrating satellites can develop orbital eccentricities comparable 
to the disc's aspect ratio. In a 3D simulation with $q=0.01$, the
flow velocity exhibits a notorious vertical component in the gap's inner edge. 
A comparison between 2D and 3D simulations reveals a slight radial offset in gap position,
resulting in a lower surface density at the perturber's orbital
radius in the 3D simulation.

\end{abstract}


\section{Introduction}
The accretion discs of active galactic nuclei (AGNs) can host stellar-mass black 
holes (BHs) \citep[e.g.,][and references therein]{tag23}. In fact,
the AGN accretion disc are commonly immersed in a nuclear star cluster and a fraction
of the initial stellar-mass BH population of this cluster can be captured by the disc
\citep{ost83,sye91,art93}. When the disc forms, about half of the initial of the stellar
BH population will be in retrograde orbit if the inner cluster has low angular momentum
\citep{sec21}. \citet{nas23} study the capture of low-mass BHs ($q\lesssim 10^{-4}$;
$q$ is defined as the ratio between the mass of the satellite BH and that of the 
central supermasive BH, often abbreviated as SMBH) 
by the accretion disc. They find that those captured BHs with an inclination larger than
a critical value $(i\sim 113^{\circ})$ increase its inclination until 
the angular momentum of the BH becomes counteraligned with the angular momentum 
of the disc (this situation corresponds to $i\sim 180^{\circ}$).

Intermediate-mass\footnote{IMBHs are defined as those 
with masses between $60$ and $10^{5}M_{\odot}$.} black holes (IMBHs) can also be
deposited into the galaxy nuclei and form a binary system with the central SMBH
\citep[e.g.,][]{por02,vol03,por06,bel10,mck12,mas14,yan19a,yan19b,sec20,szo21,dim23,pen23}.
After the binary has formed, later accretion event can rejuvenate the accretion disc
with gas having uncorrelated angular momentum \citep[e.g.,][]{nix11,ima18,imp19}.
\citet{nix11} show that the counteralignment occurs if the inclination
obeys the condition $\cos i < -J_{d}/(2J_{b})$, where $\vecJ_{d}$ and $\vecJ_{b}$
are the angular momentum of the disc and the binary, respectively. Interestingly,
they argue that the orbital shrinking of the binary is more efficient in a retrograde disc.

The interaction between the accretion disc and an orbiter in a 
(coplanar) retrograde orbit has been studied by several authors \citep{mck14,iva15,
san18,san20,sec21}. Retrograde BHs embedded in accretion discs can migrate toward 
the central SMBH and may eventually merge with it.
\citet{iva15} employ both semi-analytical 
axisymmetric models and 2D simulations to estimate the torque acting on a
retrograde satellite BH with $q\lesssim 0.02$.  \citet{san18} argue that density waves excited at distances $\lesssim H$,
where $H$ is the scaleheight of the disc, can contribute significantly to the torque.
The orbital evolution of low-mass retrograde perturbers with eccentricities 
between $0$ and $0.6$ was studied numerically in \citet{san20}. 

When the satellite is sufficiently massive, it can open a gap in the surface density
of the disc. \citet{iva15} also develop a criterion for the formation of a gap, and 
compare the radial gap profiles using axisymmetric models with 
those obtained in the simulations. 

Retrograde discs can be also found in Be/X-ray binaries.
\citet{ove24} investigate the interaction between the circumprimary disc (around
the Be star) and the companion, finding that the disc can become unstable 
to tilt perturbations near the companion's orbital radius.

In this paper, we revisit the criteria for gap formation and investigate the structure of gaps 
created by embedded non-accreting retrograde perturbers 
with $q\simeq 0.005-0.015$,  
on fixed circular orbits, using two-dimensional (2D) and three-dimensional (3D) 
simulations.
Notably, we find that the perturber does not generally reside at the bottom of the gap
but is instead located near its inner wall. Furthermore, we quantify the eccentricity of the 
gap, as an eccentric rather than circular gap could allow the perturber to
intersect it at certain points along its orbit. This interaction is expected to enhance
the BH mass accretion rate.
We also examine the sensitivity of the results to the inner boundary conditions (IBCs).
Additionally, we present some simulations where the perturber is allowed to migrate.

The paper is organized as follows. In Section \ref{sec:background} we provide the
theoretical context and present some open issues about the interaction of a
retrograde satellite embedded in the midplane of a gas disc.
Section \ref{sec:sims_general} describes the set-up and numerical aspects of the
simulations. Section \ref{sec:2D_sims}
provides the outcomes of our 2D simulations. In Section \ref{sec:3D_sims}, we discuss the results of a 3D simulation. Finally, in Section \ref{sec:conclusions}, we present our
main conclusions.

\section{Theoretical context and outstanding questions}
\label{sec:background}

We consider a thin accretion disc around a central object with mass $M_{\bullet}$.
Initially the disc is axisymmetric with surface density $\Sigma(r)\propto r^{-p}$
and gas pressure scale height $H(r)$.
The unperturbed accretion disc has aspect ratio $h(r)$ and Shakura-Sunyaev
viscosity parameter $\alpha(r)$.
The gas rotates at a velocity $\sim \Omega_{K} r$ (neglecting the gradient pressure),
where $\Omega_{K}(r)$ is the Keplerian frequency. The sound speed of the unperturbed disc is $c_{s}(r)=\Omega_{K} H(r)$.

The disc aspect ratio and the $\alpha$ parameter in the accretion discs
around SMBHs are rather uncertain.  \citet{kro99} estimates $h\simeq 0.01$ from radio-continuum emission. On the other hand, models predict that the aspect ratio of AGN accretion discs can vary by orders of magnitude with radial distance $r$. 
According to the model of \citet{sir03}, $h$ reaches a minimum value of $0.01$
at  $r\simeq 10^{3}R_{s}$, where $R_{s}$ is the Schwarzschild radius.
It then increases up to $h=0.1$ at both $r\simeq 10R_{s}$ and $r=10^{5}R_{s}$.
Meanwhile, the model of \citet{tho05} indicates that $h$ could be as small as $10^{-3}$.
Regarding the Shakura-Sunyaev $\alpha$-parameter, simulations of accretion discs suggest that it may range between $10^{-3}$ and $10^{-1}$ \citep{dav10}.

We suppose that a gravitational body (representing the satellite or perturber),
whose gravitational potential is
\begin{equation}
\Phi_{p} (\vecr,t)= -\frac{G M_{p}}{\sqrt{(\vecr-\vecr_{p})^{2}+R_{\rm soft}^{2}}}
\label{eq:Phi_p},
\end{equation}
is rotating in circular retrograde orbit  in the midplane of the disc. Here $M_{p}$ and $\vecr_{p}(t) $ are, respectively, the mass and position vector of the perturber, 
and $R_{\rm soft}$ is a softening radius. In this study, we focus on the gravitational
interaction between the disc and the perturber, while neglecting gas accretion onto the 
latter.

The relative velocity of the perturber with respect to the local gas is $2\Omega_{p} r_{p}$, 
as it moves in a retrograde circular orbit,
where $\Omega_{p}\equiv \Omega_{K}(r_{p})$.
Assuming the sound speed of the disc gas 
is constant over time (i.e. a locally isothermal disc), the Mach number of the perturber, $\mathcal{M}$, is expressed as $2\Omega_{p} r_{p}/c_{s}=2/h_{p}$, where $h_{p}$
denotes the disc aspect ratio at the perturber's location. 
Consequently,  $\mathcal{M}$ is significantly large; for instance, $\mathcal{M}=40$ when $h_{p}=0.05$.

\subsection{Condition for opening a gap}

We expect that if the accretion radius of the satellite BH, $R_{\rm acc}=2GM_{p}/(2\Omega_{p}r_{p})^{2} = qr_{p}/2$, is comparable to or larger than the disc scale height $H$, the satellite BH will have a 
strong impact on the structure of the disc, likely opening a deep gap. This 
strong shock condition occurs when $q\gtrsim q_{B}\equiv 2h$.
However, it is important to note that the corresponding strong shock condition in the prograde
case is not required to open a gap \citep[e.g.,][]{duf13,fun14,kan15}. Thus, this condition should be understood
as sufficient but not necessary.

\begin{figure}
\hskip 0.1cm
\includegraphics[angle=0,width=70mm,height=58mm]{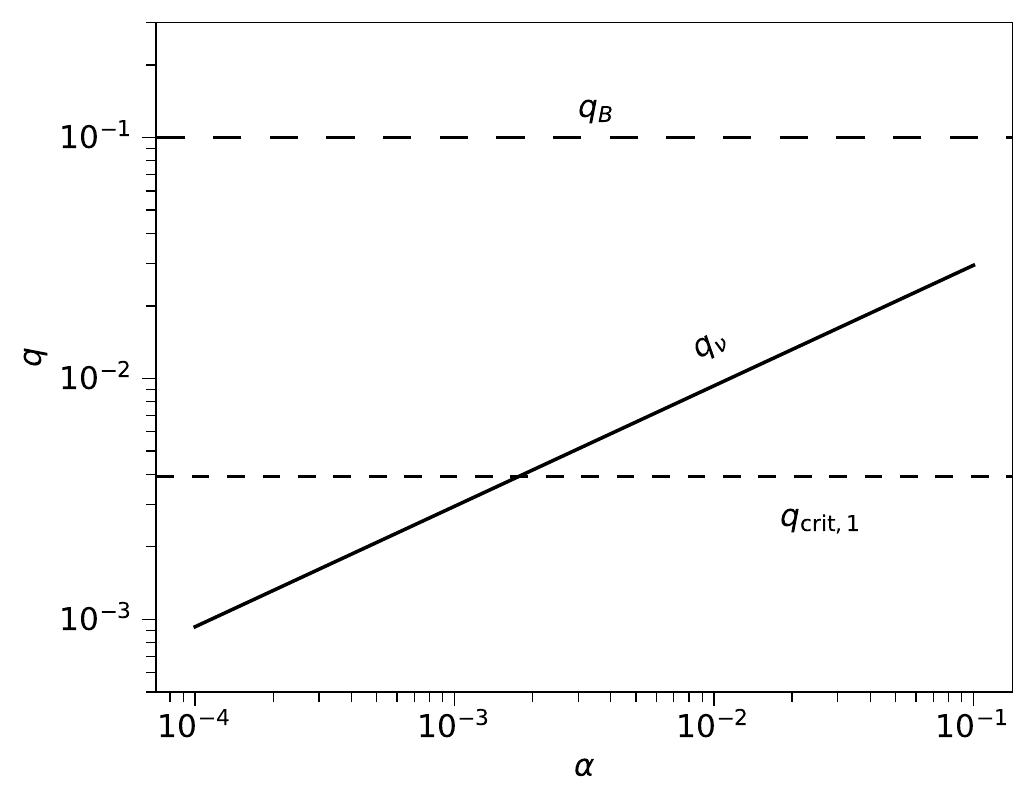}
\vskip -0.05cm
 \caption{$q_{\nu}$ (solid line), $q_{B}$ (long dashed line) and $q_{{\rm crit},1}$ (short dashed line), as a function of $\alpha$,
in a disc with $h=0.05$ and a perturber with a softening parameter $\mathcal{E}=0.6$.}
 \label{fig:q_nu}
\end{figure}

In the following, we infer the condition for the gap not to be refilled by viscosity (or
viscous condition), in the retrograde case.
The viscous torque in the accretion disc without the perturber is 
$3\pi r^{2} \nu \Sigma_{\rm un} \Omega_{K}$, where $\Sigma_{\rm un}$ is the
unperturbed surface density of the disc. Now suppose that the accretion radius  of the
perturber
$R_{\rm acc}= qr_{p}/2$ is smaller than $R_{\rm soft}$. In
that case and following the same approximation as in the prograde case
\citep[e.g.,][]{goo01,fun14,duf15}, we assume that the excitation of the wake can
be calculated in linear theory. In Appendix \ref{app:linear_formula}, we briefly 
compile the formula
for the torque in linear theory. Hence, we will assume the torque excited 
in the disc is described by Equations (\ref{eq:lambda_Gamma}) and (\ref{eq:ivanov_approx}), with the unperturbed surface density, $\Sigma_{{\rm un},p}$, replaced by the local surface density, $\Sigma_{p}$, once the gap has been carved in the disc.
The viscous condition is
\begin{equation}
3\pi r_{p}^{2} \nu \Sigma_{{\rm un},p} \Omega_{p} \leq \frac{\pi}{4}
\frac{\lambda_{\mathcal{E}} q^{2} \Sigma_{p}\Omega_{p}^{2} r_{p}^{5}}{R_{\rm soft}},
\end{equation}
or
\begin{equation}
q^{2} \geq q_{\nu}^{2}\equiv 12 \left(\frac{h_{p}\mathcal{E}}{\lambda_{\mathcal{E}}}\right) \left(\frac{\nu}{\Omega_{p} r_{p}^{2}}\right)\left(\frac{\Sigma_{{\rm un},p}}{\Sigma_{p}}\right),
\end{equation}
where $\mathcal{E}\equiv R_{\rm soft}/H$.
If we demand that the minimum depth to have a gap is $\Sigma_{p}\lesssim 0.2\Sigma_{{\rm un},p}$ then
\begin{equation}
q_{\nu}^{2}\equiv 60 \left(\frac{h_{p}\mathcal{E}}{\lambda_{\mathcal{E}}}\right) \left(\frac{\nu}{\Omega_{p} r_{p}^{2}}\right).
\label{eq:def_q_nu}
\end{equation}
In terms of the Shakura-Sunyaev viscosity parameter $\alpha$
\begin{equation}
q_{\nu}^{2}= \frac{60\mathcal{E} \alpha h_{p}^{3}}{\lambda_{\mathcal{E}}}.
\label{eq:cond_gap_formation}
\end{equation}
\renewcommand{\arraystretch}{1.35}
\begin{table}
	\centering
	\caption{Parameters of the 2D models. A gap is deep if $\left<\Sigma\right>_{\rm gap}\leq \Sigma_{{\rm un},p}/4$.
}
\label{table:params} 

 \begin{tabular}{|c|c|c|c|c|c}\hline

\multicolumn{6}{c}{Simulations with $h=0.05$}\\
 \#&  $q$ & $\nu_{-5}$ & $\mathcal{E}$ & $q/q_{\nu}$& Comments     \\ 
               
\hline 

 1 & $0.005$ & $1$ & $0.6$ & 0.85 &No deep gap  \\
 2 & $0.005$ & $4$ & $0.3$ & 0.77& No deep gap  \\
 2B & $0.005$ & $4$ & $0.6$ & 0.55& No gap  \\
 3 & $0.01$ & $1/4$ & $0.6$ &3.39 & Deep gap  \\
 4 & $0.01$ & $1$ & $0.6$ & 1.70& Deep gap  \\
 5 & $0.01$ & $4$ & $0.6$ & 0.85& Deep gap  \\
 6 & $0.015$ & $1$ & $0.6$ & 2.54& Deep gap  \\

 \hline 

\multicolumn{6}{c}{Simulation with $h=0.025$}\\ 
7 & $0.005$ & $1$ & $0.6$ & $1.20$ & Deep gap  \\
\hline 

\multicolumn{6}{c}{Simulation with $h=0.1$}\\ 
8 & $0.015$ & $1$ & $0.6$ & $1.80$ & Deep gap  \\

\hline 

\end{tabular}

\end{table}

\citet{iva15} propose a different gap-opening condition by requiring some level
of non-linearity in the interaction between the perturber and the disc. 
They argue that the sound radius $r_{c_{s}}\equiv q r_{p}/h_{p}$ should be larger 
than the most long wavelength of the waves excited by the perturber and obtain that 
a retrograde perturber opens a gap if $q>q_{{\rm crit},1} \equiv 1.57h_{p}^{2}$. 

Figure \ref{fig:q_nu} compares $q_{B}$, $q_{\nu}$ and $q_{{\rm crit},1}$ for $\mathcal{E}=0.6$ and $h_{p}=0.05$. For $\alpha\simeq 2\times
10^{-3}$  (or, equivalently, $\nu\simeq 0.5\times 10^{-5}\Omega_{p}r_{p}^{2}$),  we obtain  $q_{\nu}\simeq q_{{\rm crit},1}\simeq 4\times 10^{-3}$. However, for values
of $\alpha$ significantly larger (smaller) than $2\times 10^{-3}$, the viscous condition predicts
that a higher (lower) mass ratio is required to open a gap. 
In Section \ref{sec:gap_criterion_sim}, we use 2D simulations to discern 
whether $q_{\nu}$ or $q_{{\rm crit},1}$ is a more reliable predictor for gap formation.

Mass accretion onto the perturber could deepen the gap.
However, the simulations by \citet{iva15} indicate that gas depletion due to accretion 
has little impact on the azimuthally-averaged surface density profile of the disc.

\subsection{Does the disc reach a steady state?}
\label{sec:theoretical_prediction}

In this subsection, we explore whether a steady-state solution exits for a retrograde perturber embedded in the disc.
To do so, we adopt the approach of \citet{duf15}, who analyzed the
structure of the gap opened by a prograde perturber in the quasi-linear steady-state
regime. For
a retrograde perturber, the quasi-linear condition requires that  $R_{\rm soft}$ 
be larger than several
$R_{\rm acc}$. Under these conditions, the perturber excites linear waves that carry (negative) 
angular momentum. 
When these waves become nonlinear as they propagate through the disc, 
they deposit the angular mometum in the disc,
changing its surface density.

\begin{figure*}
\hskip -1.0cm
\includegraphics[angle=0,width=190mm,height=49mm]{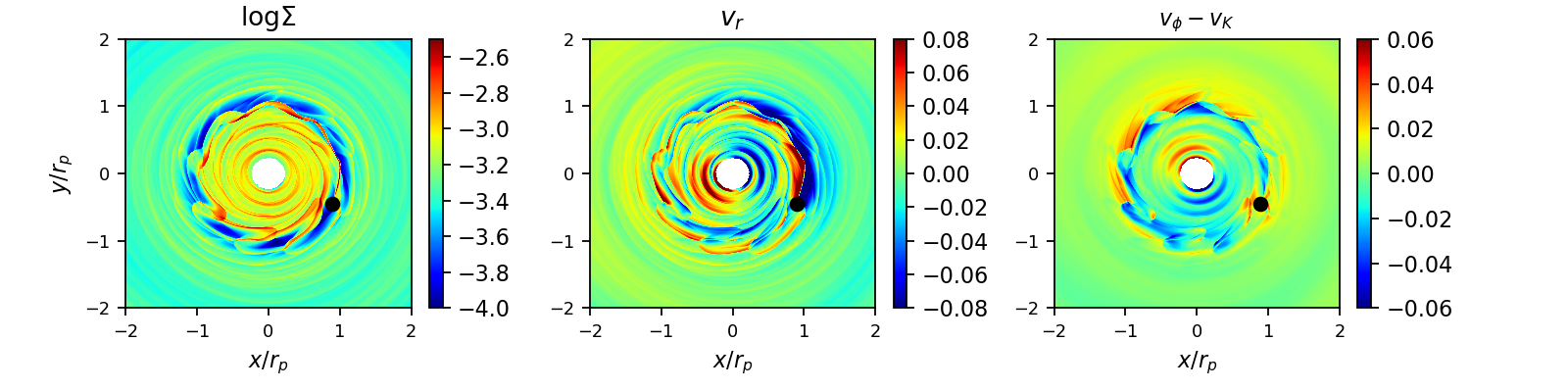}
\vskip 0.2cm
 \caption{Colour maps of the surface density (left panel), radial velocity (centre)
and $v_{\phi}-v_{K}$ (right panel; where $v_{K}$ represents the Keplerian velocity) 
in 2D simulations after $15$ perturber's orbital periods for 
model 4 ($q=0.01$ and $\nu_{-5}=1$). The perturber's position is indicated with
a black dot. It moves clockwise, whereas the disc rotates counterclockwise. 
}
 \label{fig:maps_2D_15orb}
\end{figure*}

If a steady state solution exists, the mass flow rate through the disc $\dot{M}$,
and the angular momentum flow $F_{J}$, should be constant with $r$. Hence 
\begin{equation}
\dot{M}=-2\pi r \Sigma v_{r}={\rm const}
\end{equation}
and 
\begin{equation}
F_{J}=-\dot{M} r^{2} \Omega_{K} +3\pi \Sigma \nu r^{2} \Omega_{K} + \Phi_{w}(r)={\rm const},
\end{equation}
where $\Phi_{w}(r)$ is the angular momentum flux transported by the waves excited
by the perturber. We consider the solution $F_{J}=0$, which implies 
\begin{equation}
-\dot{M} r^{2} \Omega_{K} +3\pi \Sigma \nu r^{2} \Omega_{K} + \Phi_{w}(r) =0.
\label{eq:duffell0}
\end{equation}
In the absence of the perturber ($\Phi_{w}=0$), the equation above implies
$\dot{M}=3\pi \nu \Sigma_{\rm un}(r)$. This mass flow is expected to be maintained in
the presence of the perturber; Equation (\ref{eq:duffell0}) transforms into:
\begin{equation}
-3\pi \nu \Sigma_{\rm un} r^{2} \Omega_{K} +3\pi \Sigma \nu r^{2} \Omega_{K} + \Phi_{w}(r) =0.
\label{eq:duffell}
\end{equation}

Using Equation (\ref{eq:lambda_Gamma}), we estimate the angular momentum
flux $\Phi_{w}(r)$ at $r>r_{p}$ and at $r<r_{p}$, denoted by $\Phi_{w}^{+}$ and $\Phi_{w}^{-}$, respectively, as 
\begin{equation}
\Phi_{w}^{\pm}(r) \simeq \mp\frac{1}{2} \lambda_{\mathcal{E}} \Gamma_{0} f(r),
\end{equation}
where $f(r)>0$ is a radial function to be specified.
If we assume that the excitation torque occurs at a neighbourhood of the
perturber, the function $f(r)$ has its maximum at $r=r_{p}$ and satisfies that $f(r_{p})=1$.
Evaluating Equation (\ref{eq:duffell}) at $r=r_{p}$, we obtain
\begin{equation}
\frac{\Sigma_{p}^{\pm}}{\Sigma_{{\rm un},p}}= \left(1\mp\frac{\lambda_{\mathcal{E}}
q^{2} \Omega_{p} r_{p}^{3}}{24\nu R_{\rm soft}}\right)^{-1} = 
\left(1\mp \frac{5 q^{2}}{2q_{\nu}^{2}}\right)^{-1}.
\label{eq:unun}
\end{equation}
In the last equality we have used Eq. (\ref{eq:def_q_nu}). 
This solution is physically unacceptable because, aside from the surface density 
jump at $r=r_{p}$, $\Sigma_{p}^{+}$ becomes negative for $q>\sqrt{2/5}q_{\nu}$. Therefore, no global
steady state exists. Nor does a steady state exist at $r>r_{p}$, assuming, as we did, 
that the excitation of the waves occurs in a small region around $r=r_{p}$.
This indicates that the problem is time-dependent, and the long-term solution may 
depend on the chosen boundary conditions (BCs).

\section{Hydrodynamical simulations: code and setup}
\label{sec:sims_general}
We simulate the evolution of a gaseous disc using the publicly available code FARGO3D
\citep{ben16,ben19} in polar coordinates $(r,\phi)$ for the 2D simulations
and spherical coordinates $(r, \theta,\phi)$ for the 3D simulations..
We use a frame where the central mass $M_{\bullet}$ 
is at rest at the origin. Initially the disc is axisymmetric with
surface density $\Sigma_{\rm un}(r) = \Sigma_{0}(r/R_{0})^{-1/2}$ and aspect ratio 
$h$ constant over radius. The evolution of the disc
is locally isothermal, that is, the sound speed $c_{s}(r)$ is constant over time. 
We also include a kinematic viscosity $\nu$ constant over the domain.

\begin{figure*}
\hskip -1.0cm
\includegraphics[angle=0,width=194mm,height=54mm]{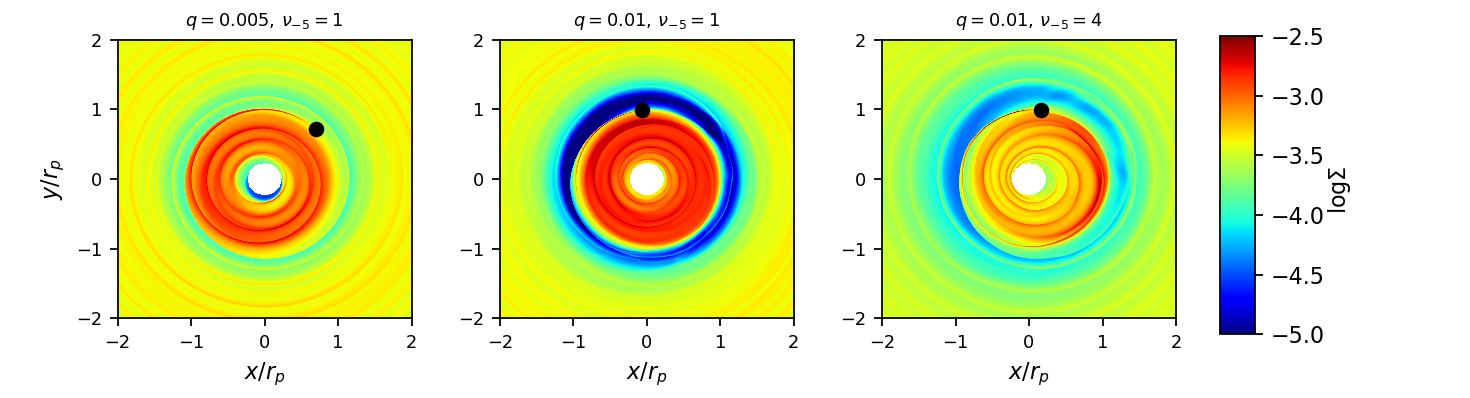}
\vskip -0.0cm
 \caption{Colour maps of the surface density in 2D simulations at $750$ orbits for models
1, 4 and 5 (from left to right). 
The body moves clockwise, whereas the disc rotates counterclockwise. 
}
 \label{fig:maps_2D}
\end{figure*}

\begin{figure}
\hskip 0.1cm
\includegraphics[angle=0,width=77mm,height=63mm]{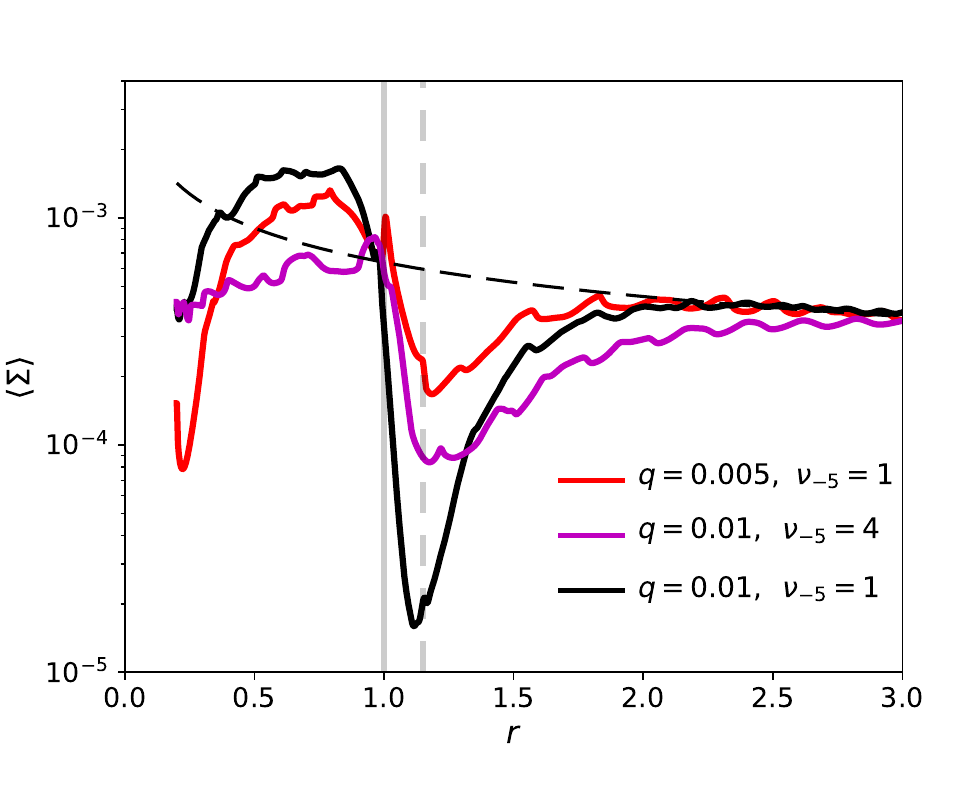}
\vskip -0.05cm
 \caption{Azimuthally averaged surface density for the three snapshots shown in Figure
 \ref{fig:maps_2D}. The long-dashed curve represents the initial surface density, while
the vertical lines indicate $r=1$ (solid) and $r=1.15$ (dashed)
for reference.} 
 \label{fig:Sigma_R_3models}
 \end{figure}

At $t=0$, we insert the perturber on a retrograde circular orbit with radius $r_{p}=1$.
Unless otherwise stated, the orbit is fixed to be circular. 
The total gravitational potential is given by
\begin{equation}
\Phi (\vecr) = -\frac{GM_{\bullet}} {r}+\Phi_{p}(\vecr)+\Phi_{I}(\vecr),
\end{equation}
where $\Phi_{p}(\vecr)$ is the potential created by the perturber, as given in 
Eq. (\ref{eq:Phi_p}), and $\Phi_{I}(\vecr)$ is the indirect term due to the shift of 
the barycentre in the presence of the perturber
\begin{equation}
\Phi_{I}(\vecr)=\frac{GM_{p}}{r_{p}^{3}} \vecr\cdot \vecr_{p}.
\end{equation}
We include neither mass accretion by the perturber, nor the gas self-gravity, nor
the indirect term due to the acceleration felt by the central body from the disc gas.

\begin{figure}
\hskip -0.3cm
\includegraphics[angle=0,width=94mm,height=44mm]{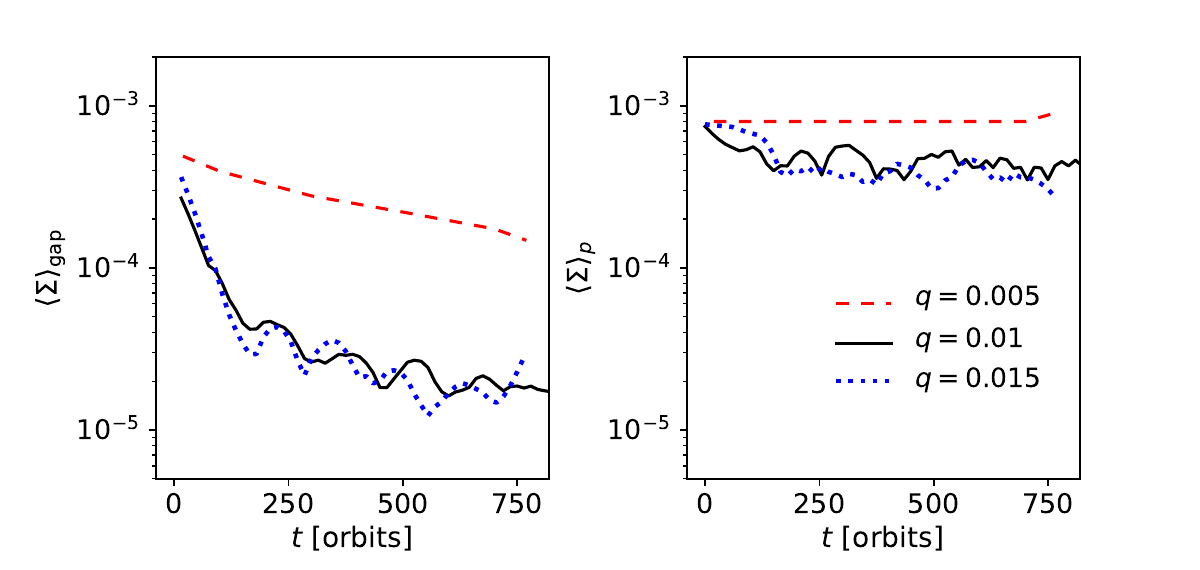}
\vskip -0.05cm
 \caption{$\left<\Sigma\right>_{\rm gap}$ (left column) and $\left<\Sigma\right>_{p}$ (right column) for $\nu_{-5}=1$, and
varying $q$.  To avoid cluttering the figure, the values are time averaged over one orbital
revolution and taken regularly every $15$ orbits. 
 }
 \label{fig:Sigma_diffq_open}
\end{figure}

\begin{figure}
\hskip -0.3cm
\includegraphics[angle=0,width=94mm,height=44mm]{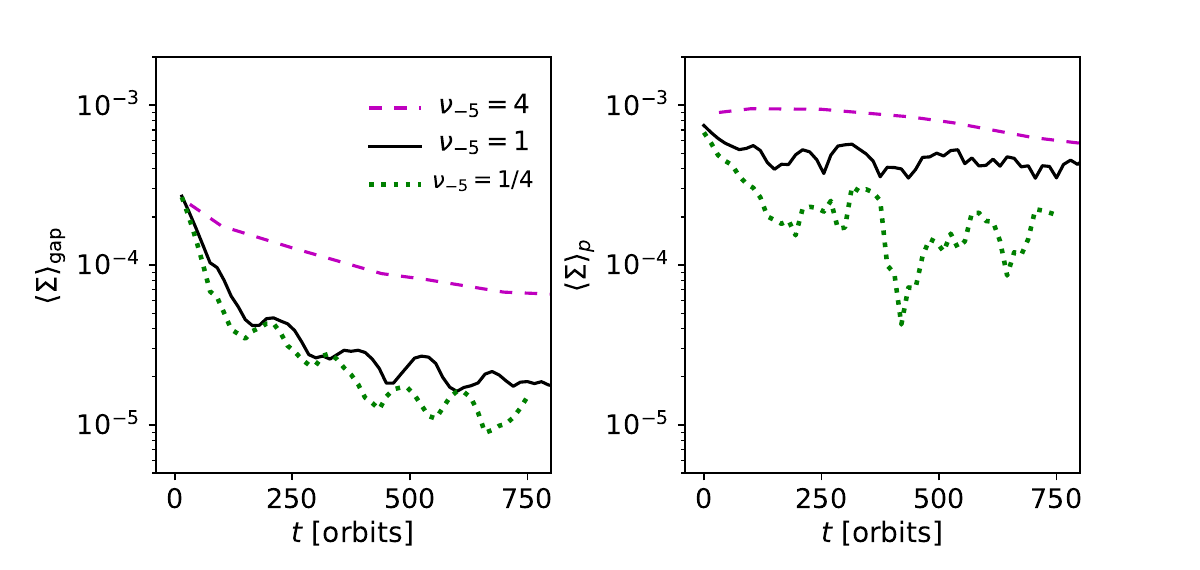}
\vskip -0.05cm
 \caption{$\left<\Sigma\right>_{\rm gap}$ (left column) and $\left<\Sigma\right>_{p}$ (right column) for $q=0.01$ and
 different viscosities. As in Fig. \ref{fig:Sigma_diffq_open}, the values were taken at $15$-orbit intervals and time averaged over one perturber's orbit. 
 }
 \label{fig:Sigma_diffnu_open}
\end{figure}

\begin{figure}
\hskip -0.0cm
\includegraphics[angle=0,width=80mm,height=60mm]{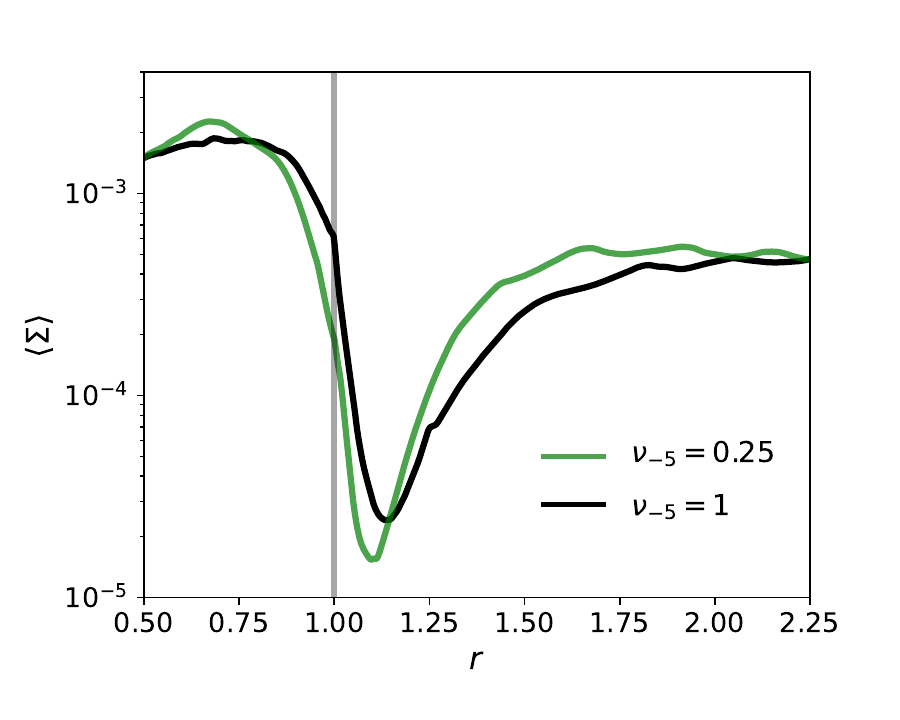}
\vskip -0.05cm
 \caption{$\left<\Sigma\right>$, averaged over $100$ orbits (from $650$ to $750$
 orbits), for $q=0.01$ and two values of the viscosity $\nu$ (models $3$ and $4$). 
The vertical line indicates the position of the perturber.
 }
 \label{fig:Sigma_nu025_nu01}
\end{figure}

\begin{figure}
\hskip -0.0cm
\includegraphics[angle=0,width=80mm,height=100mm]{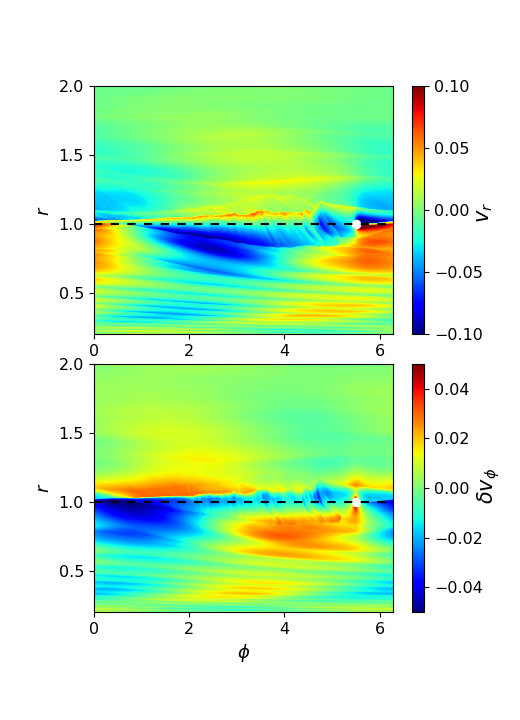}
\vskip -0.05cm
 \caption{Radial velocity (top panel) and azimuthal velocity perturbations 
$v_{\phi}-v_{K}$ (bottom panel) for model $7$ after $50$ orbits.  
The white dot marks the position of the perturber, which moves from right to left.
The dashed black line delineates the boundary between the inner and the outer disc.
 }
 \label{fig:vrad_vphi_50orb}
\end{figure}

\begin{figure}
\hskip 0.1cm
\includegraphics[angle=0,width=80mm,height=62mm]{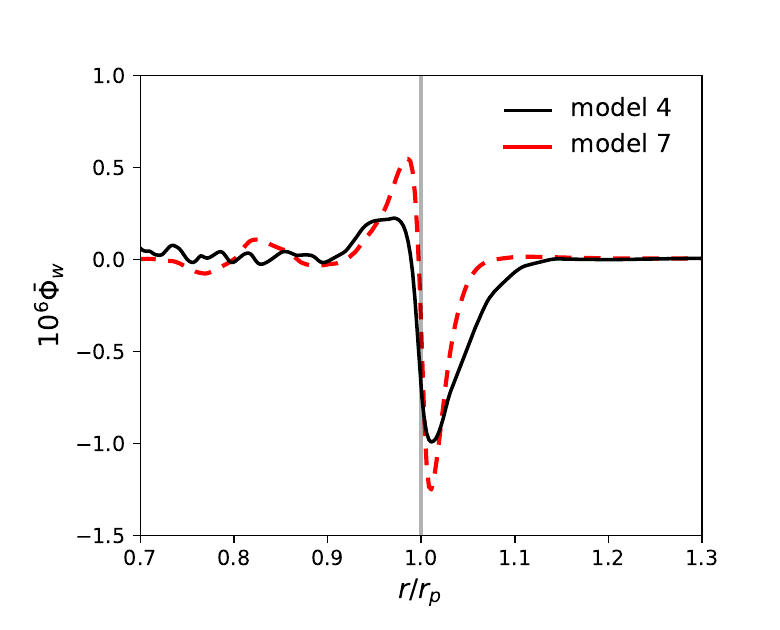}
\vskip -0.05cm
 \caption{Angular momentum flux carried by propagating density disturbances, 
$\bar{\Phi}_{w}$ (see Eq. \ref{eq:AM_flux}), at $t=750$ orbits.}
 \label{fig:AM_flux}
\end{figure}

\begin{figure}
\hskip 0.1cm
\includegraphics[angle=0,width=80mm,height=138mm]{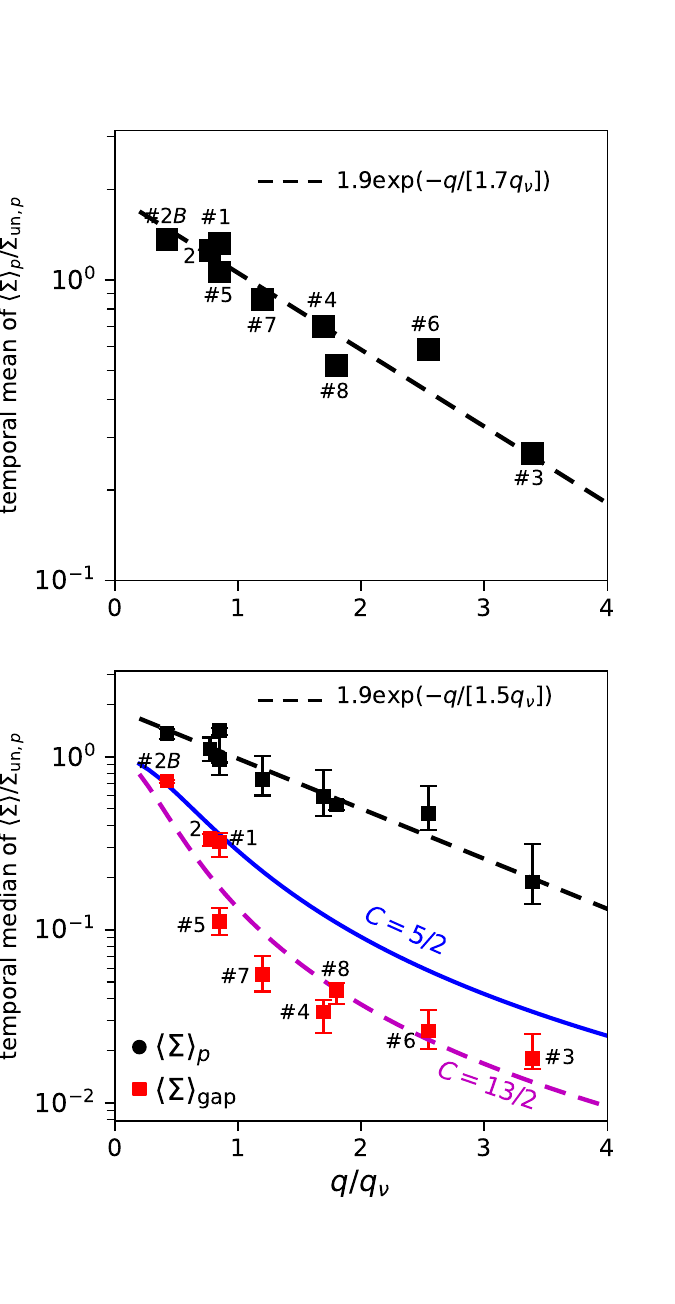}
\vskip -0.05cm
 \caption{Top panel: Temporal mean value of $\left<\Sigma\right>_{p}$,
normalized to $\Sigma_{{\rm un},p}$, for the models given in Table \ref{table:params}. 
The dashed black line represents the function $\left<\Sigma\right>_{p}/\Sigma_{{\rm un},p}= 1.9\exp(-q/[1.9q_{\nu}])$.
Bottom panel: Temporal median values
of $\left<\Sigma\right>_{p}$ (black dots) and 
$\left<\Sigma\right>_{\rm gap}$
(red squares), normalized to $\Sigma_{{\rm un},p}$, for the same models.
The error bars in the bottom panel indicate the $16$th and $84$th percentile. 
The mean and median values were taken between $500$ to $750$ orbits.
The blue and magneta curves represent the simple formula given in Equation (\ref{eq:simple_C}) with $C=5/2$ and $C=13/2$, respectively. The dashed black line
in the bottom panel
represents the function $\left<\Sigma\right>_{p}/\Sigma_{{\rm un},p}= 1.9\exp(-q/[1.6q_{\nu}])$.}
 \label{fig:test_prediction}
\end{figure}

\begin{figure}
\hskip 0.1cm
\includegraphics[angle=0,width=80mm,height=68mm]{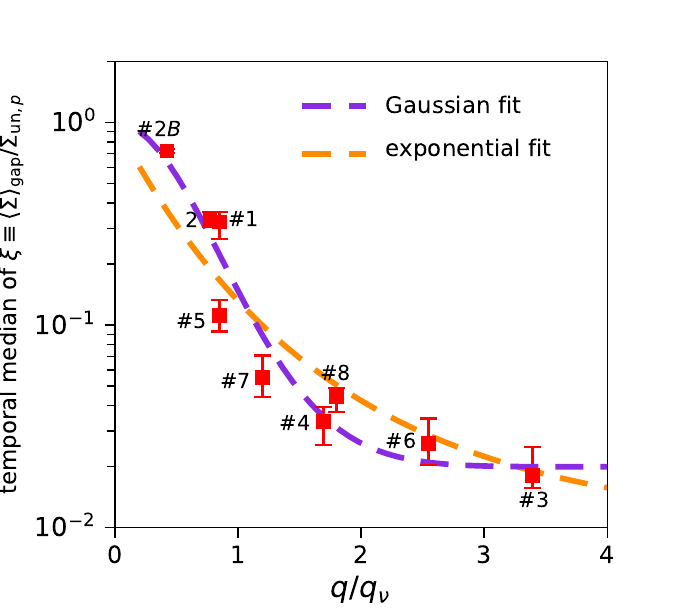}
\vskip -0.05cm
 \caption{Other fits to the temporal median values of 
$\xi\equiv \left<\Sigma\right>_{\rm gap}/\Sigma_{{\rm un},p}$
(red squares). The violet line
represents the function $\log\xi= 1.7(\exp\{-0.67[q/q_{\nu}]^{2}\}-1)$, whereas the orange represents the function $\log\xi= 2(\exp\{-0.58[q/q_{\nu}]\}-1)$.}
 \label{fig:exp_Gauss_fits}
\end{figure}

\section{2D simulations}
\label{sec:2D_sims}
In all our 2D simulations, 
the inner and outer boundaries are set at $r=0.2$ and $r=4$, respectively.
We use a grid with $1200$ cells in both the radial and azimuthal directions. 
The radial grid is logarithmic. We apply damping conditions \citep{deV06}
in the outer boundary
to prevent reflections of density waves. At the inner edge, we 
use outflow BCs, that is, we allow mass outflow through the inner
boundary (but not mass inflow is permitted)\footnote{With the exception of model 8, where damping IBCs are applied to mitigate the excessive depletion of material in the inner disc caused by outflow IBCs.}. However, we 
explore the impact of using other IBCs (damping and rigid BCs) in Section \ref{sec:IBCs}.

We consider three values for the perturber-to-central mass ratio: $q=0.005, 0.01$
and $0.015$; and three values for the disc viscosity: $\nu_{-5}=0.25, 1$ and $4$, where
$\nu_{-5}\equiv \nu/(10^{-5}\Omega_{p} r_{p}^{2})$. In most simulations, 
$h$ is set to $0.05$, and the softening radius of the perturber's gravitational
potential is taken as $R_{\rm soft}=0.6H$ (corresponding to $\mathcal{E}=0.6$). 
Nevertheless, given the significant variation of $h$ with $r$ 
expected in AGN accretion discs
(see Section \ref{sec:background}), we have also considered other values of $h$.
A summary of the models is given in Table \ref{table:params}.

\subsection{Azimuthally averaged surface density}
\label{sec:surface_density}
At early stages, the flow pattern is far from axisymmetric (see Fig. \ref{fig:maps_2D_15orb}).
The radial velocity map clearly reveals the gravitational deflection of streamlines
behind the perturber. Even as early as $t=15$ perturber's orbital periods, the gas ahead
of the perturber is highly disturbed, exhibiting alternating overdense and 
underdense regions. 
Within $r<0.6r_{p}$, blue and red arcs are visible in both $v_{r}$ and $v_{\phi}-v_{K}$
(here $v_{K}$ is the Keplerian velocity),
indicating that the disc eccentricity is being excited. The behaviour in this early phase 
of evolution differs from the prograde case, where gaps form smoothly in viscous
discs \citep[e.g.,][]{kle99}.
 
Figure \ref{fig:maps_2D} shows the disc surface density for models 1, 4 and 5,
after $750$ orbits. In all three cases,
the perturber is situated in the edge of the gap. This feature is clearly seen in the 
azimuthal-averaged surface density profile, which we denote
as $\left<\Sigma\right>$ (see Figure \ref{fig:Sigma_R_3models}).
For $q=0.01$ and $\nu_{-5}=1$, 
the bottom of the gap is located at $r\simeq 1.15$.
In terms of $H$, it is radially shifted outward by a distance $3H$ from the perturber's
orbital radius.
As a consequence, the azimuthally-averaged surface density at $r=r_{p}$,
$\left<\Sigma\right>_{p}$,
is about $20$ times larger than $\left<\Sigma\right>_{\rm gap}$, the azimuthally 
averaged surface density at the bottom of the gap. Indeed, $\left<\Sigma\right>_{p}\simeq
\Sigma_{{\rm un},p}\equiv \Sigma_{\rm un}(r_{p})$.
This is in contrast with the gap opened by a prograde perturber, where 
the perturber is located at gap's bottom.

Figure \ref{fig:Sigma_diffq_open} shows $\left<\Sigma\right>_{\rm gap}$ and 
$\left<\Sigma\right>_{p}$,
as a function of time, for $\nu_{-5}=1$ and three different values of $q$.
The notorious difference between the values of $\left<\Sigma\right>_{\rm gap}$ 
and  $\left<\Sigma\right>_{p}$ reflects the fact that
the radius of the gap's minimum does not coincide with the perturber's orbital radius.
$\left<\Sigma\right>_{\rm gap}$ presents a decreasing trend with time.
It is noteworthy that $\left<\Sigma\right>_{\rm gap}$ decreases
a factor of $\sim 10$ as $q$ increases from $0.005$ to $0.01$, but shows
only a slight variation when $q$ changes from $0.01$ to $0.015$.

For $q=0.01-0.015$, $\left<\Sigma\right>_{p}$ presents temporal variations 
about a mean value of $\sim 4\times 10^{-4}$, which is only a factor $2/3$ smaller 
than its value at $t=0$.

Figure \ref{fig:Sigma_diffnu_open} shows $\left<\Sigma\right>_{\rm gap}$ and 
$\left<\Sigma\right>_{p}$ for $q=0.01$ and different disc
viscosities. It is clear that the gap is shallower for the highest
value of the viscosity ($\nu_{-5}=4$). 
Interestingly, $\left<\Sigma\right>_{p}$ becomes smaller and exhibits larger temporal
variations as viscosity decreases. 
One could think that $\left<\Sigma\right>_{p}$ would decrease in discs with low
viscosity due to the widening of the gap.
However, our simulations indicate that, in contrast to the prograde case 
\citep[e.g.,][]{kan16},
the gap width in the retrograde configuration increases with viscosity. The value
$\left<\Sigma\right>_{p}$ is smaller for $\nu_{-5}=0.25$, because
the radial shift between the gap minimum and the perturber's orbital radius
decreases for lower values of $\nu_{-5}$ (see Figure \ref{fig:Sigma_nu025_nu01}).

The indirect term $\Phi_{I}$ takes into account the reflex motion of the central star 
around the centre of mass. We have explored the role of the indirect term by comparing 
the azimuthally averaged surface density
$\left<\Sigma\right>$, $\left<\Sigma\right>_{\rm gap}(t)$ and 
$\left<\Sigma\right>_{p}(t)$,  for a simulation with and without the indirect term 
$\Phi_{I}$ in the gravitational potential. 
We found that these quantities are not affected, in a statistical sense, by the indirect term.

\subsection{Why is the perturber offset from the minimum of the gap?}
The result that a non-migrating perturber does not lie at the bottom but
at edge of the gap is a bit unexpected. In fact, we have applied an imposed (external)
torque to the disc (without any gravitational perturber), $|\Lambda_{\rm imp}(r)|
\propto \exp(-\lambda[r-r_{0}]^{2})$, and confirmed that the resulting gap has its
minimum at $r=r_{0}$.

The simplest hypothesis for the radial shift between the gap and the perturber is
the distinct dynamical responses of the inner ($r<r_{p}$)
and outer $r>r_{p}$) disc regions (see Figures \ref{fig:maps_2D_15orb} and \ref{fig:vrad_vphi_50orb}). The asymmetry between the inner and outer parts 
of the disc arises in the non-linear regime
as soon as the perturber reaches its own wake.
This asymmetrical behaviour likely results in a reduction of
the deposition of (negative) angular momentum within the inner disc.

As an indication of this potential reduction in angular momentum deposition within the 
inner disc, we have computed
\begin{equation}
\Phi_{w}(r) = r^{2} \int_{0}^{2\pi} (v_{\phi}-\left<v_{\phi}\right>) (v_{r}-\left<v_{r}\right>)\,
\Sigma\,d\phi,
\label{eq:AM_flux}
\end{equation}
which serves as a proxy for the angular momentum flux carried by the propagating density
disturbances.
Figure \ref{fig:AM_flux} presents $\bar{\Phi}_{w}(r)$, where the overbar denotes an
average over one orbit of the perturber, calculated after it has completed $750$ orbits, for models 4 and 7. In both cases, $\bar{\Phi}_{w}$ is far from being
antisymmetric with respect to $r=r_{p}$. For both models, density waves excited
in the outer disc transport (negative) angular momentum outward. Interestingly,
although waves are also excited in the inner disc, they carry significantly less 
angular moment over one perturber's orbital period, particularly in model 4.

Therefore, we attribute the shift between the perturber and the centre of the gap to
the unequal transfer of angular momentum between the perturber and the inner 
disc compared to the perturber and the outer disc. We further suggest that models
excluding this effect \citep[e.g.][]{iva15} are likely to predict gaps that are excessively deep.

\subsection{Gap formation criterion}
\label{sec:gap_criterion_sim}
 
In this section, we study how $\left<\Sigma\right>_{\rm gap}$ and $\left<\Sigma\right>_{p}$ scale with the parameter $q/q_{\nu}$. The upper panel of 
Figure \ref{fig:test_prediction} shows the time- and azimuthally-averaged surface density at $r=r_{p}$, in units of $\Sigma_{{\rm un},p}$, as a function
of $q/q_{\nu}$, for the simulations in Table \ref{table:params}. The average over time was done from $500$ to $750$ orbits. The lower panel shows the median values of $\left<\Sigma\right>_{p}$
and $\left<\Sigma\right>_{\rm gap}$ in the same time series.
All these quantities decay with $q/q_{\nu}$ (albeit with some scatter). 
The functional dependence can be fitted reasonably well with
\begin{equation}
\frac{\left<\Sigma\right>}{\Sigma_{{\rm un},p}}=\left(1+\frac{Cq^{2}}{q_{\nu}^{2}}\right)^{-1}.
\label{eq:simple_C}
\end{equation}
For the mean and median values of $\left<\Sigma\right>_{p}$, we find $C=0.59$ and $C=2/3$, respectively. For the median value of $\left<\Sigma\right>_{\rm gap}$,
$C\simeq 13/2$. 
Albeit with some scatter, $(q/q_{\nu})^{2}$ can be considered an ordering parameter. 
The coefficient of $C=5/2$ predicted in Section \ref{sec:theoretical_prediction}  overestimates the value of $\left<\Sigma\right>_{\rm gap}$ and underestimates the value of $\left<\Sigma\right>_{p}$. 

We explored whether alternative functional forms 
provide a better fit than Equation (\ref{eq:simple_C}). Define
$\xi\equiv \left<\Sigma\right>_{\rm gap}/\Sigma_{{\rm un},p}$.
Figure \ref{fig:exp_Gauss_fits} 
reveals that the exponential dependence of $\log\xi$ on $q/q_{\nu}$ provides a
comparable fit to Eq. (\ref{eq:simple_C}), whereas the Gaussian fit yields a
sligthly closer agreement.  However, the
flat behaviour of $\log\xi$ at large $q/q_{\nu}$ values predicted by the Gaussian
curve appears to be physically unrealistic.

Note that $\left<\Sigma\right>_{p}\gtrsim \Sigma_{{\rm un},p}$
for $q\leq q_{\nu}$. In the particular case of model 1, the density enhancement in 
the wake behind the perturber overbalances the density reduction due to the 
angular momentum transfer between the perturber and the disc (see 
Fig. \ref{fig:Sigma_R_3models}).
On the other hand, it can be seen
that $q\gtrsim 1.2q_{\nu}$ is a sufficient (but not necessary) condition for the formation of a deep gap.

In all the models from $1$ to $6$ in Table \ref{table:params}, we have 
taken $h=0.05$. Consequently, $q_{{\rm crit},1}=0.004$ for these models.  
Since all the simulations in Figure \ref{fig:test_prediction}
satisfy $q >q_{{\rm crit},1}$ (see Table \ref{table:params}), all of them should present a gap according to this criterion. However, in models 1, 2 and 2B, 
where $q=0.005$, the perturber cannot open a deep gap.

\begin{figure}
\hskip -0.0cm
\includegraphics[angle=0,width=92mm,height=99mm]{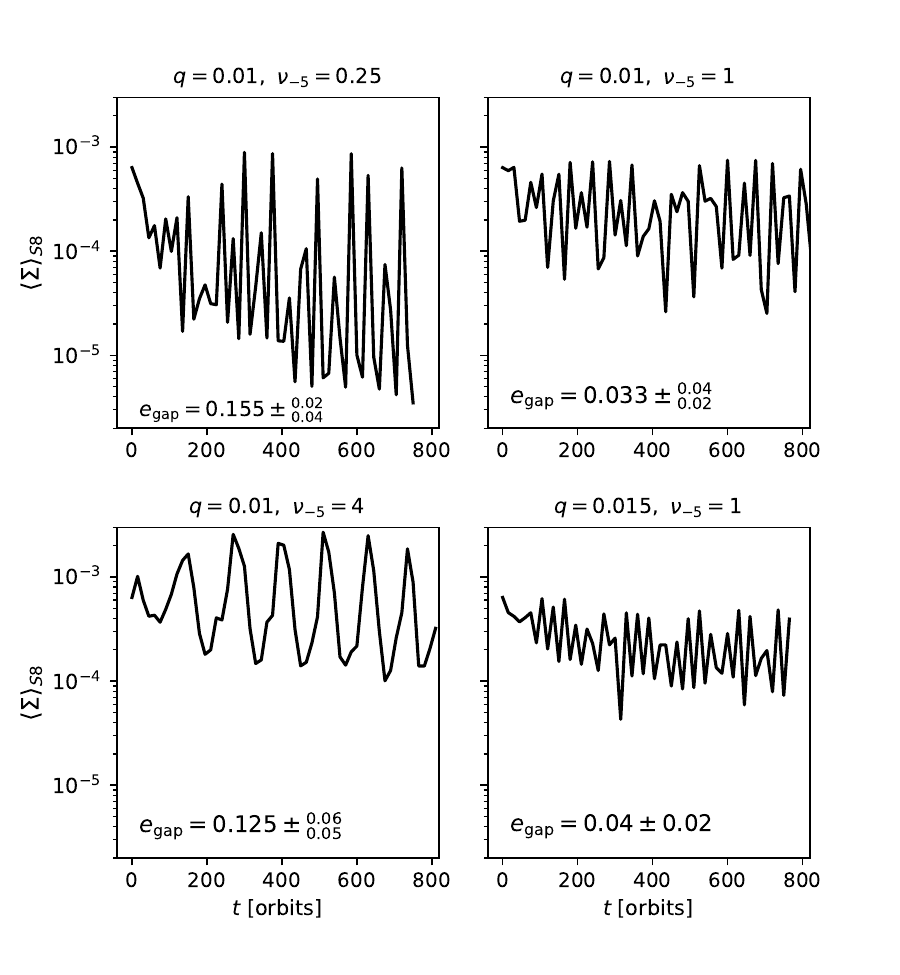}
\vskip -0.05cm
 \caption{$\left<\Sigma\right>_{S8}$ at $r=r_{p}$ as a function of time for runs 3, 4,
5 and 6. To keep the figure uncluttered, the values were taken at $15$-orbit intervals.
The periodic oscillations of $\left<\Sigma\right>_{S8}$ in model 5 result from an aliasing effect, as the
sampling rate is low compared to the true frequency of the oscillations.
 }
 \label{fig:Sigma_front}
\end{figure}

\begin{figure}
\hskip -0.04cm
\includegraphics[angle=0,width=80mm,height=158mm]{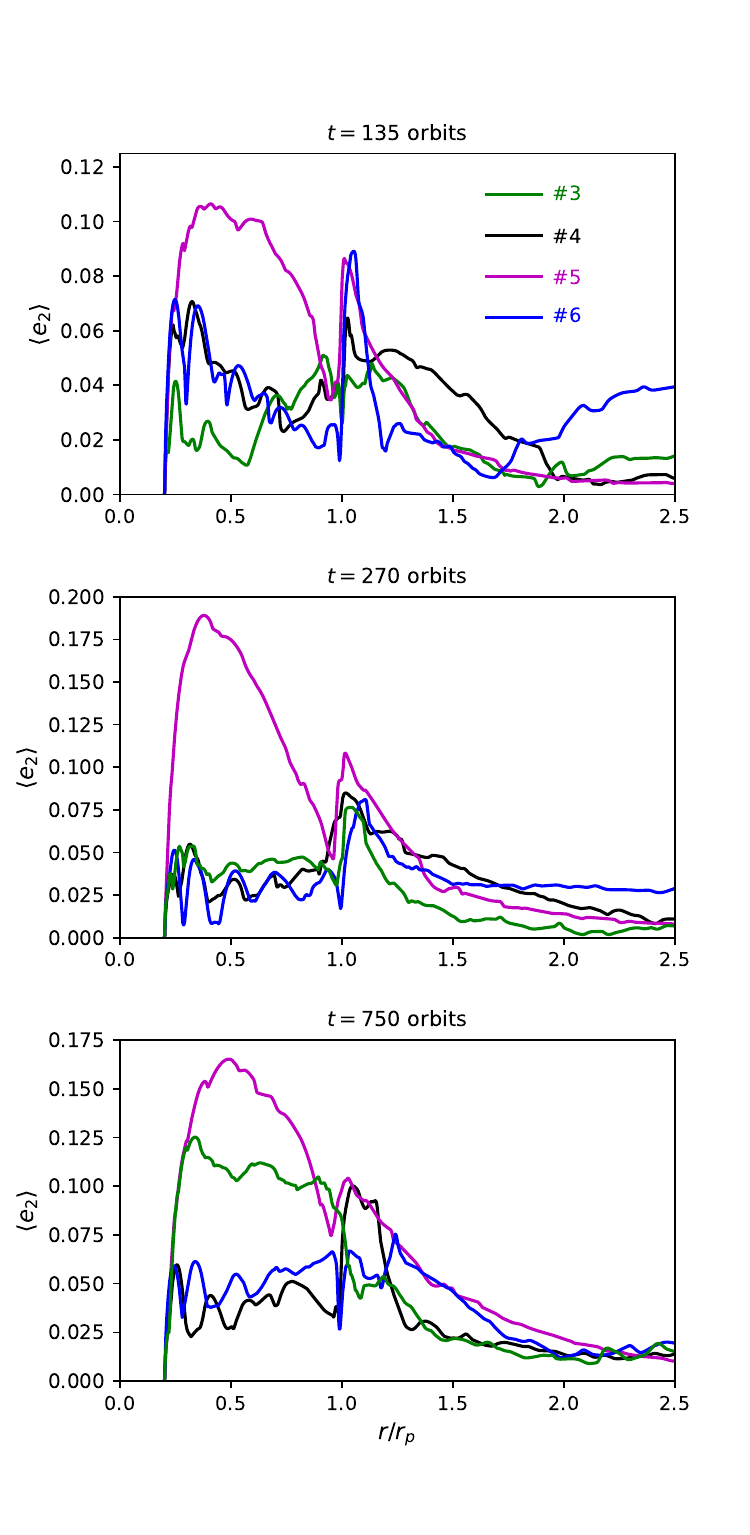}
\vskip -0.0cm
 \caption{Disc eccentricity, $\left<e_{2}\right>$,
 at three differentn times: $135$ orbits (upper panel),  $270$ orbits (central panel) and 
$750$ orbits (lower panel), for models 3, 4, 5 and 6.
}
 \label{fig:disc_eccentricity_openIBCs}
\end{figure}

\subsection{The fluctuating flow around the perturber }
\label{sec:eccentricity}
In Section \ref{sec:surface_density}, we have shown that the perturber is located in the 
inner wall
of the gap, i.e. in a region with a steep radial density gradient. Therefore, both axisymmetric
oscillatory modes and non-axisymmetric radial motions in the disc,
if exist, are expected to produce
a very fluctuating flow pattern around the body.

In order to quantify this flow variability, we have divided the disc into eight
equally-sized azimuth sectors. We denote S1 to the azimuth sector just behind
the perturber (it extends in azimuth from $\phi=\phi_{p}$ to $\phi_{p}+\pi/4$,
where $\phi_{p}$ is the perturber's azimuthal angle), whereas S8 is the sector just
ahead of the perturber. More specifically, we define the surface density averaged over
sector S$j$ as
\begin{equation}
\left<\Sigma\right>_{Sj}= \frac{4}{\pi}\int_{\phi_{p}+(j-1)\beta}^{\phi_{p}+j\beta}
\Sigma(r,\phi,t) \,\,d\phi,
\end{equation}
with $\beta=\pi/4$.  Since S8 is the sector just at the front of the perturber,
the mass accretion rate onto the perturber (if it were an accretor, which is not the case
in our simulations) is more directly related to 
$\left<\Sigma\right>_{S8}$ at $r=r_{p}$ than to $\left<\Sigma\right>_{p}$.

Figure \ref{fig:Sigma_front} shows 
$\left<\Sigma\right>_{S8}$ at $r=r_{p}$, taken every $15$ orbits,
for four representative models.
It is apparent that $\left<\Sigma\right>_{S8}$ at $r=r_{p}$ is more fluctuating than
$\left<\Sigma\right>_{p}$ (compare the right-hand panel of Fig. \ref{fig:Sigma_diffnu_open} 
and Fig. \ref{fig:Sigma_front}). This is expected because $\left<\Sigma\right>_{p}$ in Figs.
\ref{fig:Sigma_diffq_open} and \ref{fig:Sigma_diffnu_open} involves an averaging process over
one orbit of the perturber. Without this averaging, $\left<\Sigma\right>_{p}$ would
exhibit variations over one orbital time due to the presence of axisymmetric 
oscillatory modes in the disc.

For the models under consideration,
the temporal mean value of $\left<\Sigma\right>_{p}$
is a factor of $\sim 1.5$ larger than $\left<\Sigma\right>_{S8}$ at $r=r_{p}$.
The reason is that, unlike $\left<\Sigma\right>_{S8}$,
$\left<\Sigma\right>_{p}$ includes the (overdense) wake trailing the perturber.

In the model with $q=0.01$ and $\nu_{-5}=1/4$, $\left<\Sigma\right>_{S8}$ at $r=r_{p}$
shows the largest variations, spanning approximately two orders of magnitude. Among the
four models, the model with $q=0.015$ displays the smallest variations in 
$\left<\Sigma\right>_{S8}$, with a variation 
factor of $\sim 6$, compared to a factor of $\sim 15$ for the $q=0.01$ model with the same viscosity.

\subsection{Gap and disc eccentricities}

Given the strong radial density gradient at the gap's edge, dynamical instabilities, 
such as the Rayleigh instability, are likely to excite the gap's eccentricity, which may
contribute to produce the observed fluctuations in $\left<\Sigma\right>_{S8}$ at $r=r_{p}$. 
Therefore, evaluating the eccentricity of the inner edge of the gap is of particular 
interest. Furthermore, modelling the interaction with axisymmetric models (as in 
Section \ref{sec:background}) is only justified if the eccentricity of the inner edge
of the gap, $e_{\rm gap}$, 
is small enough that $e_{\rm gap}r_{p}$ remains less than the width of the gap.

To compute $e_{\rm gap}$, we follow this procedure. First, we define the edges of the gap using two isocontours corresponding to a surface density of 
$\Sigma_{{\rm un},p}/2$. At a given time $t$, the inner edge is represented by a polar equation $r(\theta)$. We then search for the value of $\theta$ at which $r$ reaches its maximum, denoted as 
$\theta_{\rm max}(t)$. The gap eccentricity is calculated using the following formula:
\begin{equation}
e_{\rm gap}(t)\equiv \frac{ r(\theta_{\rm max})-r(\theta_{\rm max}+\pi)} {r(\theta_{\rm max})+r(\theta_{\rm max}+\pi)}.
\label{eq:egap}
\end{equation} 
Note that this definition of the gap eccentricity is meaningful only for deep gaps,
i.e. those having $\left<\Sigma\right>_{p}\leq \Sigma_{{\rm un},p}/4$. 

The complex interaction between the perturber and the edge
of the gap induces temporal variations in $e_{\rm gap}$. Therefore, 
in Table \ref{table:egap} and in each panel of Figure \ref{fig:Sigma_front},
we report the median value of $e_{\rm gap}$ along with the $16$th and
$84$th percentiles, calculated over the time series from $500$ to $750$ orbits.
We see that $e_{\rm gap}$ may acquire large values ($\sim 0.18$) in runs 3 and 5.
Nevertheless, the effect  of eccentric modes on $\left<\Sigma\right>_{S8}$ at $r=r_{p}$
depends not only on $e_{\rm gap}$ but also on the density gradient.
For instance, in simulation 5, although $e_{\rm gap}$
is large, the variations in $\left<\Sigma\right>_{S8}$ remain moderate
due to the relatively shallow density radial gradient compared to the other simulations.

In run 4, the gap remains relatively circular. However, as viscosity is either decreased or 
increased, the gap becomes more eccentric. This unexpected result that $e_{\rm gap}$
is not a monotonic function of viscosity, suggests a spurious excitation of $e_{\rm gap}$
in the inner boundary in run 5. It is therefore important to track the growth of the
eccentricity at various distances in the disc.

To do so, we have computed the ``disc eccentricity'' at different times. 
Some authors refer to the disc eccentricity as the azimuthally
averaged value of $e_{\rm disc}$, which is defined as 
\begin{equation}
e_{\rm disc}(r,\phi,t) \equiv \sqrt{\left(\frac{v_{\phi}^{2}}{v_{K}^{2}}-1\right)^{2}
+\frac{v_{r}^{2}v_{\phi}^{2}}{v_{K}^{4}}},
\end{equation}
where $v_{K}(r) =\sqrt{GM_{\bullet}/r}$ is the Keplerian velocity \citep{kle06,tan22}.
The azimuthal averaged of $e_{\rm disc}$, $\left<e_{\rm disc}\right>$, measures how much
a ring of the disc separates from Keplerian rotation. \citet{tan22} suggest that
$e_{2}'\equiv |v_{r}v_{\phi}|/v_{K}^{2}$ is a better estimator of how a ring separates
from circular rotation. Still, an axisymmetric disc with a radial flow has
a non-null $e_{2}'$. Therefore, we expect that $\left<e_{2}\right>$ with
\begin{equation}
    e_{2}\equiv \frac{|(v_{r}-\left<v_{r}\right>)v_{\phi}|}{v_{K}^{2}}
\end{equation}
is a better proxy of asymmetry.

Figure \ref{fig:disc_eccentricity_openIBCs} shows
$\left<e_{2}\right>$ versus $r$, at three different times (at $t=135$, $270$ and $750$ orbits), in models $3$, $4$, $5$ and $6$. Note that, in general, 
$\left<e_{2}\right>$ at $r=r_{p}$ is smaller than $e_{\rm gap}$. 
Thus, $\left<e_{2}\right>$ should not be interpreted as the ellipticity of the 
surface density isocontours, but as a measurement of disc asymmetries.
In most of the cases, $\left<e_{2}\right>$ exhibits 
a local maximum at $r\simeq 1.15$, indicating the signature of the wake excited
by the perturber. At $t=750$ orbits, $\left<e_{2}\right>$ at $r=r_{p}$ lies between
$0.04$ and $0.1$.

In model 5, we observe that $\left<e_{2}\right>$ reaches its peak around 
$r\simeq 0.4r_{p}$ in the three panels,
suggesting that, at $t=135$ orbits and beyond, the excitation of the disc eccentricity 
due to the interaction of wakes with the inner boundary may contribute to enhance 
the gap eccentricity. In models 4 and 6, the inner computational
boundary does not appear to significantly impact the gap eccentricity, at least up
to $t=750$ orbits. Conversely, in model 3, spurious interactions between the
wakes and the inner boundary seem to drive an increase in the gap eccentricity beyond 
$\sim 360$ orbits. 

In summary,  from $500$ to $750$ orbits, $\left<e_{2}\right>$ 
may be dominated by artificial boundary effects in models 3 and 5. In contrast, 
the impact of the inner boundary on $\left<e_{2}\right>$ remains small
in models 4 and 6, where $\left<e_{2}\right>$ is $\sim 0.07$ at
the $84\%$ of the distribution.
In the following subsection, we examine how the structure of the gap changes when
other IBCs are applied.

\begin{figure}
\hskip 0.1cm
\includegraphics[angle=0,width=86mm,height=73mm]{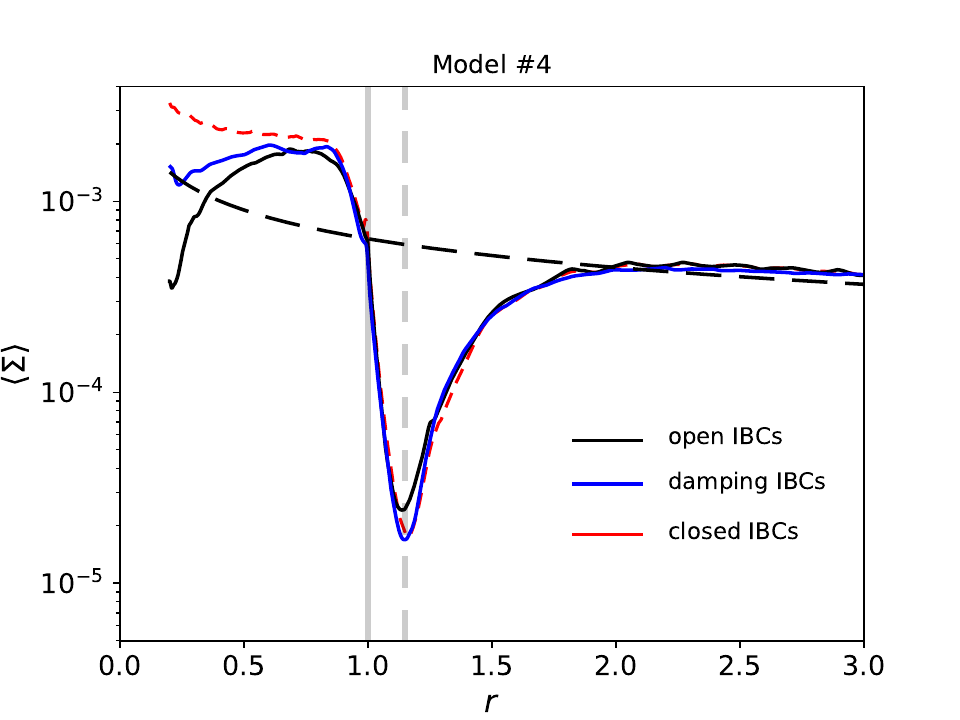}
\vskip -0.05cm
 \caption{Surface density averaged over both azimuth and time, between $650$ and $750$ orbits, using different IBCs:
outflow (solid line), damping (blue line) and rigid (red line) IBCs. 
The initial surface density profile is also
 shown with a long dashed line. The vertical solid grey line indicates $r=1$ (where the perturber is located) and 
 the dashed grey line indicates $r=1.15$ (which
 is approximately the location of the bottom of the gap). In all cases, $q=0.01$ and
$\nu_{-5}=1$.}
 \label{fig:Sigma_R_diffBC}
\end{figure}

\begin{figure*}
\hskip -1.4cm
\includegraphics[angle=0,width=194mm,height=45mm]{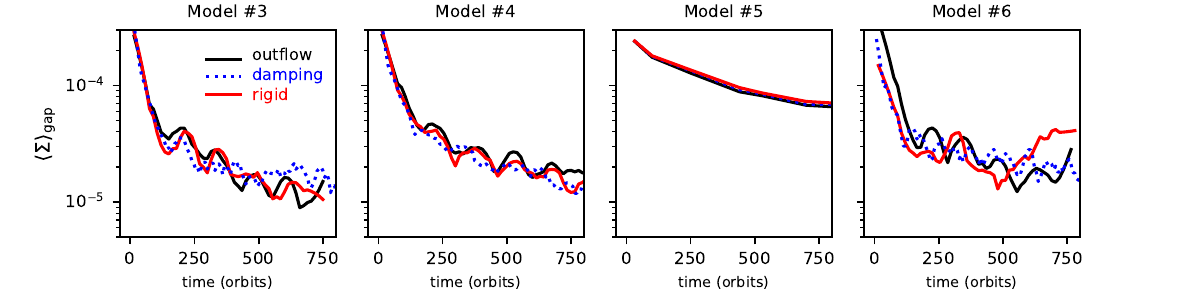}
\vskip -0.0cm
 \caption{Comparison of $\left<\Sigma\right>_{\rm gap}$  for
 different IBCs: outflow (black lines), damping (blue lines) and rigid (red lines) in
models 3, 4, 5 and 6. As in Figures \ref{fig:Sigma_diffq_open} and \ref{fig:Sigma_diffnu_open}, 
$\left<\Sigma\right>_{\rm gap}$ was computed every $15$ orbits of the perturber
and time averaged over one orbit. }
 \label{fig:Sigma_min_diffBCs}
\end{figure*}

\begin{figure*}
\hskip -1.4cm
\includegraphics[angle=0,width=194mm,height=45mm]{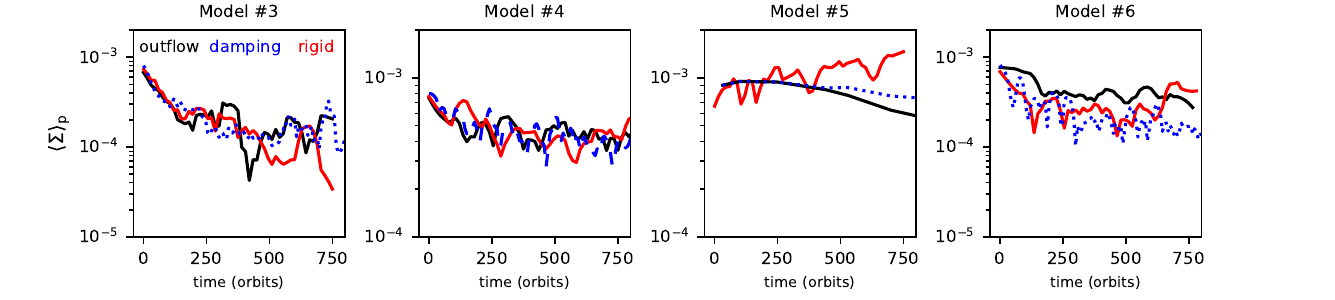}
\vskip -0.0cm
 \caption{Similar to Figure \ref{fig:Sigma_min_diffBCs} but for $\left<\Sigma\right>_{p}$.
}
 \label{fig:Sigma_r1_diffBCs}
\end{figure*}

\begin{figure*}
\centering
\includegraphics[angle=0,width=199mm,height=54mm]{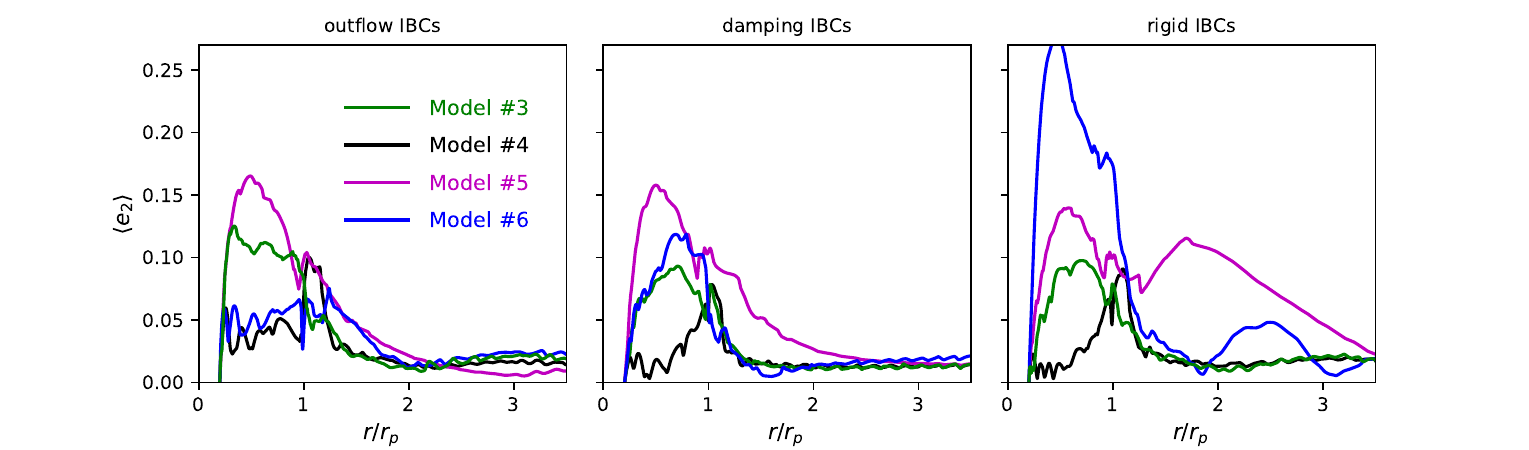}
 \caption{$\left<e_{2}\right>$ at $750$ orbits,
 for outflow (left panel), damping (middle panel) and rigid (right panel) IBCs, for
models 3, 4, 5 and 6.
}
 \label{fig:e2_diffBCs}
\end{figure*}

\subsection{The impact of the IBCs}
\label{sec:IBCs}
In the prograde case, damping BCs, as described in \citet{deV06}, are the most commonly applied in both the inner and outer boundaries of the computational domain. 
Prior to the introduction of damping BCs, 
some authors implemented outflow IBCs \citep[e.g.,][]{kle99,sne01,dan03,hos07}, whereas other authors chose
rigid IBCs \citep[e.g.,][]{lub99,kle01,bat03,nel03}.
For times shorter than the viscosity timescale and with boundaries sufficiently
distant from the perturber, the torque on a prograde perturber and the gap 
structure are largely unaffected by the BCs. This is because spiral waves are 
excited at the Lindblad and corotation resonances, and the torque is dominated by resonances at distances of
$\sim h$ from the perturber \citep{gol80}. 
Since there are no resonances between the orbiter and the 
disc in the circular retrograde case, the disc response, and consequently the 
torque acting on the orbiter, are expected to be more sensitive to the IBC than in
the prograde case.

\renewcommand{\arraystretch}{1.35}
\begin{table}
	\centering
	\caption{Median values, $16$th, and $84$th percentiles of $e_{\rm gap}$ (as given in  Eq. \ref{eq:egap}) over the time series from $500$ to $750$ orbits for the three
types of IBCs in 2D simulations. 
}
\label{table:egap} 
 \begin{tabular}{|c|c|c|c|}\hline
 \#&     $e_{\rm gap}$  & $e_{\rm gap}$  & $e_{\rm gap}$  \\ 
  & outflow IBCs  & damping IBCs & rigid IBCs \\
               
\hline 

 3 &   $0.155\substack{+0.02\\-0.04}$ & $0.12\substack{+0.02\\-0.04} $ & $0.10\substack{+0.05\\-0.01}$\\
 4 &   $0.033\substack{+0.04\\-0.02}$ &  $0.035\substack{+0.015\\-0.03}$ &
$0.043\substack{+0.01\\-0.02}$\\
 5 &  $0.125\substack{+0.06\\-0.05}$ & $0.135\substack{+0.04\\-0.05}$ &
$0.095\substack{+0.04\\-0.02}$\\
 6 &  $0.04\pm 0.02$ & $0.16\substack{+0.02\\-0.03}$ & $0.25\substack{+0.17\\-0.05}$\\
7  &   $0.065\substack{+0.01\\-0.03}$ & $0.060\substack{+0.02\\-0.025}$ & $0.065\substack{+0.02\\-0.035}$\\
8 &   $--$ & $0.027\substack{+0.035\\-0.02}$ & $--$\\

 \hline 

\end{tabular}  

\end{table}

In this subsection, we explore how the results depend on the adopted IBC in 2D simulations.
We compare the results using three types of IBCs: outflow, damping and rigid. 
Figure \ref{fig:Sigma_R_diffBC} shows the azimuthally-averaged 
(and time averaged) surface density for model 4.
We see that beyond $r=0.8r_{p}$, the
radial profiles of $\left<\Sigma\right>$ are similar for the three IBCs.

Figures \ref{fig:Sigma_min_diffBCs} and \ref{fig:Sigma_r1_diffBCs} show $\left<\Sigma\right>_{\rm gap}$ and $\left<\Sigma\right>_{p}$, respectively,
as a function of time, for different IBCs. 
In all the models shown in Fig. \ref{fig:Sigma_min_diffBCs},
$\left<\Sigma\right>_{\rm gap}$ exhibits a decreasing trend over time, except
for model 6 with rigid IBCs. 
A comparison of results for outflow and
damping IBCs reveals that both predict similar values for 
$\left<\Sigma\right>_{\rm gap}$ and $\left<\Sigma\right>_{p}$, except
in model 6, where the damping IBC predicts a 
$\left<\Sigma\right>_{p}$ value lower by a factor of $2$ (see the fourth panel in Fig. \ref{fig:Sigma_r1_diffBCs}).

Figure \ref{fig:e2_diffBCs} shows $\left<e_{2}\right>$ at 
$750$ orbits, for the same models as depicted in Figures  \ref{fig:Sigma_min_diffBCs} and 
\ref{fig:Sigma_r1_diffBCs}. Among the three IBCs, model 4 exhibits the lowest
eccentricity excitation at $r<r_{p}$. In model 5, the disc eccentricity exhibits significant excitation at the boundary 
across all three types of IBCs. In this model, $\left<e_{2}\right>$ at $750$
orbits is similar for both outflow and damping IBCs. In model 6, 
the disc eccentricity is less excited with outflow IBCs, whereas in model 3,
outflow IBCs result in the highest $\left<e_{2}\right>$ values at $r<r_{p}$.

The gap eccentricity, $e_{\rm gap}$, is also expected to be sensitive to the IBCs.
Table \ref{table:egap} provides the median values of $e_{\rm gap}$ over 
the time series from $500$ to $750$ orbits for the three different types of IBCs.
For model 3,  $e_{\rm gap}$ is sligthly higher when using outflow IBCs.
In model 4, $e_{\rm gap}$ is not very different across the three types of IBCs.
In model 5, the lowest $e_{\rm gap}$ is obtained for rigid IBCs, while in model 6, 
$e_{\rm gap}$ is significantly lower with outflow IBCs.
Consequently, no specific IBC
minimizes disc eccentricity across all four models.

\begin{figure}
\hskip 0.1cm
\includegraphics[angle=0,width=81mm,height=64mm]{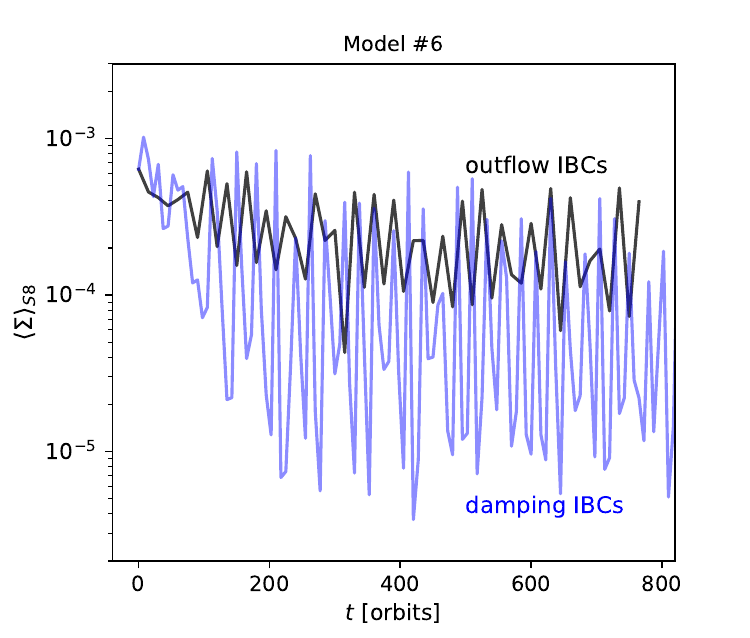}
\vskip -0.05cm
 \caption{$\left<\Sigma\right>_{S8}$ at $r=r_{p}$ as a funtion of time, using outflow IBCs
(black line) and damping IBCs (blue line), in model 6 ($q=0.015$ and
$\nu_{-5}=1$). }
 \label{fig:Sigma_front_diffBC}
\end{figure}

Fig. \ref{fig:Sigma_front_diffBC} shows $\left<\Sigma\right>_{S8}$ at $r=r_{p}$
in model 6,
using outflow and damping IBCs. The figure clearly demonstrates the sensitivity of
$\left<\Sigma\right>_{S8}$ to $e_{\rm gap}$. In fact, 
$\left<\Sigma\right>_{S8}$ displays greater temporal variations
with damping IBCs as $e_{\rm gap}$ is significantly
larger under these IBCs.

In the previous and current subsections, we have kept the perturber's orbit fixed and 
focused on the disc eccentricity. However, the perturber's eccentricity can also grow.
In Appendix \ref{sec:migrating}, we present the radial migration and the evolution of the
eccentricity of the perturber when migration is allowed.

\begin{figure*}
\includegraphics[angle=0,width=200mm,height=61mm]{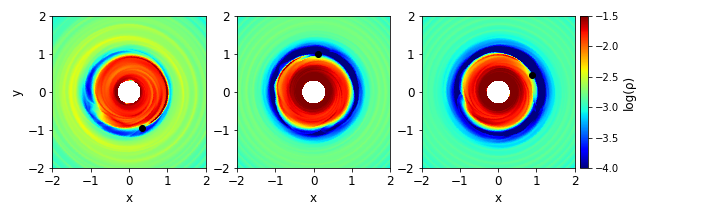}
\includegraphics[angle=0,width=200mm,height=61mm]{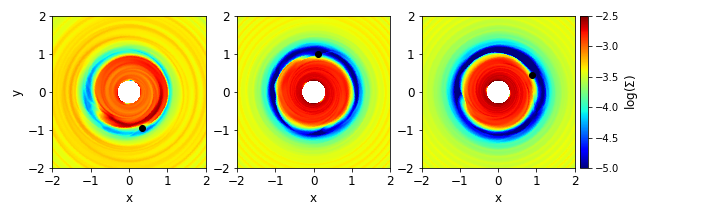}
\vskip -0.0cm
 \caption{Volume density at $z=0$ (upper panels) and surface density (lower
panels) at $60$, $690$ and $1500$ orbits (from left to right).
The instantaneous position of the perturber is marked with a black circle. The perturber
moves clockwise. }
 \label{fig:maps_faceon_3D}
\end{figure*}

\begin{figure*}
\includegraphics[angle=0,width=199mm,height=105mm]{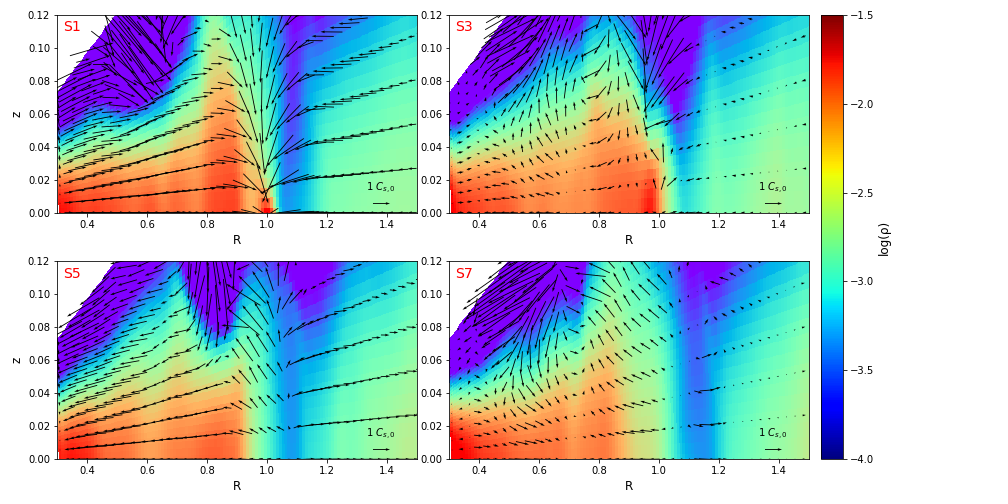}
\hskip -2.3cm
\vskip -0.75cm
 \caption{Meridional flow fields in four different sectors at $t=60$ orbits. Background color
represents volume gas density. }
 \label{fig:meridional_60orb}
\end{figure*}

\begin{figure*}
\includegraphics[angle=0,width=199mm,height=105mm]{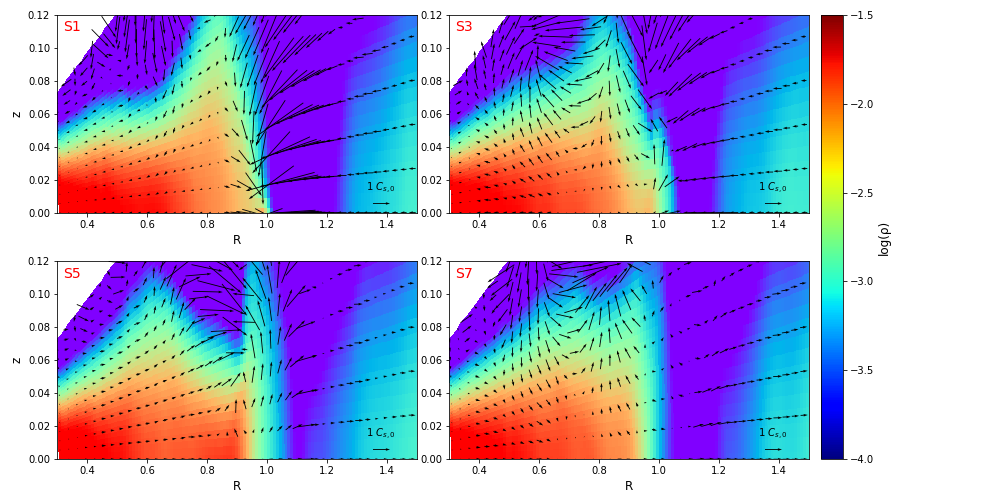}
\hskip -2.3cm
\vskip -0.75cm
 \caption{Same as Figure \ref{fig:meridional_60orb} but at $t=1500$ orbits.}
 \label{fig:meridional_1500orb}
\end{figure*}

\begin{figure}
\hskip 0.1cm
\includegraphics[angle=0,width=86mm,height=73mm]{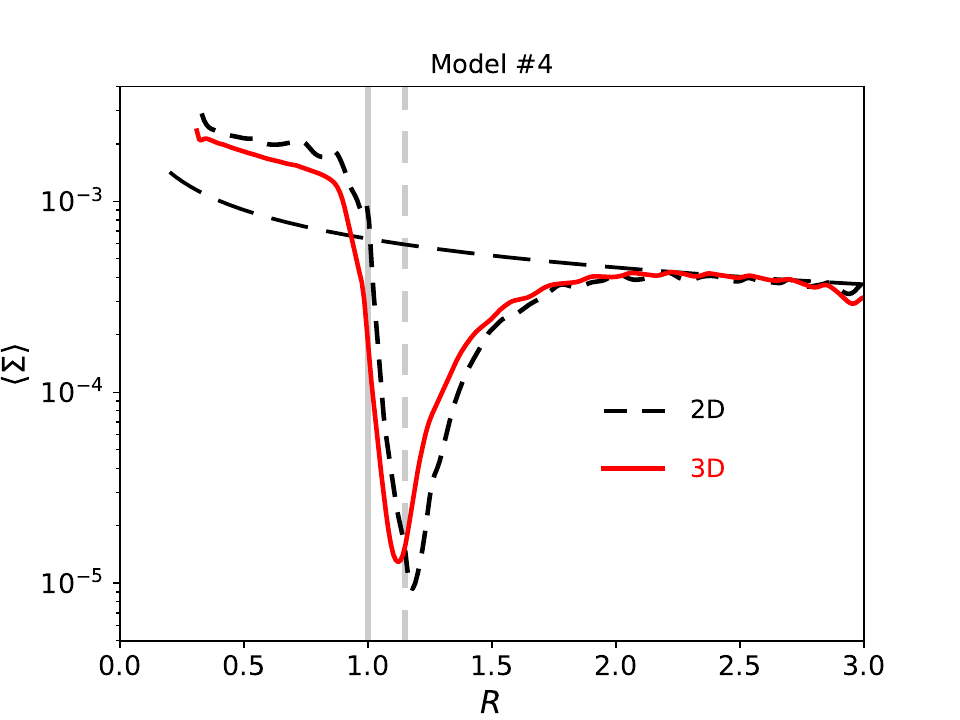}
\vskip -0.05cm
 \caption{Azimuthally averaged surface density after $690$ orbits 
in a 2D simulation (short dashed line) and in a 3D simulation (solid red line).
The long dashed line corresponds to the initial surface density. For reference,
the vertical solid grey line indicates $R=1$  and 
 the dashed grey line indicates $R=1.15$.
 }
 \label{fig:Sigma_R_3D}
\end{figure}

\section{3D simulation}
\label{sec:3D_sims}
Our 2D simulations indicate that the satellite is embedded in the inner edge of the gap,
a region with a large density gradient.
To study the 3D structure of the gap and, in particular, the flow at the edges of the gap,
we have performed one 3D simulation with the same parameters of $q$, $\nu$ and 
$\mathcal{E}$ as in model 4. We simulate only the upper half of the
disc. The inner boundary is located at $r=0.3$, the outer boundary at $r=3$, and
the upper boundary at a latitude angle of $13.75^{\circ}$, which corresponds to 
$z_{\rm max}=4.8H$, where $H$ is the vertical disc scaleheight.
The number of cells are $n_{r}=276$, $n_{\phi}=576$, and $n_{\theta}=40$. 
Since the softening radius $R_{\rm soft}$ is $0.6H_{p}$, we have $3$ zones per
$R_{\rm soft}$ in the radial and azimuthal directions, and $5$ zones per $R_{\rm soft}$
in the vertical direction. 
The boundaries are rigid in the inner and outer radii.
At the upper cap of the domain, we apply open BCs.
As in our 2D simulations, the disc is initialized with a 
power-law radial profile $\Sigma \propto R^{-0.5}$ and aspect ratio $h=0.05$.

\begin{figure*}
\includegraphics[angle=0,width=179mm,height=60mm]{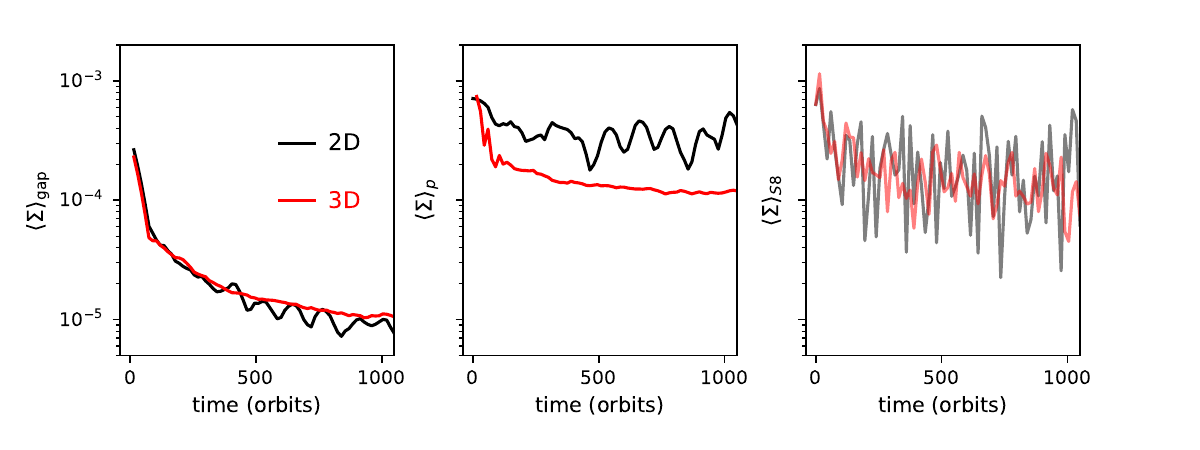}
\hskip -2.3cm
\vskip -0.75cm
 \caption{$\left<\Sigma\right>_{\rm gap}$ (left panel), 
$\left<\Sigma\right>_{p}$ (central panel) and 
$\left<\Sigma\right>_{S8}$ at $r=r_{p}$ (right panel) versus time for 2D (black line) 
and 3D simulations (red line) in model 4. Both simulations have the same domain 
size and number of cells in the radial and azimuthal directions. }
 \label{fig:Sigma_min_r1_S8}
\end{figure*}

Figure \ref{fig:maps_faceon_3D} shows the volume density at the midplane
of the disc ($z=0$) and the surface density (i.e. the vertically integrated volume density)
at three different times. We observe that the maps of volume density at $z=0$
and surface density are highly similar, suggesting that the dynamics 
of the disc's bulk mass is largely cylindrical. A very thin tail behind the perturber is 
visible in the three maps of the volume density at $z=0$, and it is less pronounced or 
not visible in the integrated surface density.

After $60$ orbits, the gap is still incipient and non-axisymmetric. We can see some gas plumes in the forming-gap.
After $690$ orbits, the gap is more deep and becomes more axisymmetric, although
some corrugations in the inner edge of the gap are visible. At $1500$ orbits, although 
the gap 
appears slightly wider,  some corrugations along the inner edge are still present.
Similar to the 2D simulations, the orbiter is positioned at  inner rim of the gap,
as observed in both the surface density and volume density maps.

Figures \ref{fig:meridional_60orb} and \ref{fig:meridional_1500orb} show the density 
and velocity vectors along $R$-$z$ slices, averaged over different sectors (S1, S3, S5
and S7), at $t=60$ orbits and at $t=1500$ orbits, respectively. Between $R\simeq 0.9$ 
and $R=1.1$, the flow velocity exhibits a notorious vertical component.
Additionally, the velocity field varies markedly across different sectors. For example, 
in many locations, the velocity arrows in sectors S1 and S5 point in opposite directions.
Unlike the prograde case \citep[see Fig. 4 in][]{fun16},   
the gas dynamics at the inner edge differ entirely from those at the outer
edge.

Let us first consider the sector S1. Recall that S1 is the sector just behind the perturber. 
In both Figs. \ref{fig:meridional_60orb} and \ref{fig:meridional_1500orb}, 
within the range $0.9\leq R\leq 1.1$, the velocity is supersonic and the vectors 
point toward the perturber, creating a region of compression. 
The velocity in this region is induced by the gravitational pull by the perturber.
A tiny density enhancement at $(R,z)=(1,0)$  
associated with the tail excited by the perturbed can be seen in S1 of Fig. \ref{fig:meridional_60orb}.
After the passage of the perturber, the high pressure generated by the vertical
compression drives gas upward in S5 and S7 and, consequently, 
the inner ``wall'' of the gap (i.e. the material between $R=0.8$ and $R\simeq 1$) 
expands in the vertical direction.

In sectors S1 and S3, the flow within the gap and the outer disc ($R\gtrsim 1.2$) moves 
radially inward. Along the gap, the flow accelerates until it collides supersonically (with Mach number $3$-$4$) with the inner wall. In contrast,
in sectors S5 and S7, the gas in the gap and in the outer disc 
(at $R>1.1$) flows radially outwards. This variation in the radial velocity direction 
across different sectors is attributed to the eccentricity of the outer edge of the gap.
As a result of this outward motion, 
the gap exhibits a marginally greater width in S7 relative to S5.

In Figure \ref{fig:Sigma_R_3D}, we plot $\left<\Sigma\right>$ at $t=690$ orbits,
for both the 2D and 3D simulations. 
In the 2D simulation, we used the same domain extension and number of zones
in both the radial and azimuthal directions as in the 3D simulation. 
The gap appears slightly more pronounced in the 2D case. Although the shape of the 
gap is similar in both simulations, they are slightly offset radially. This radial offset, 
though small, results in a lower value of $\left<\Sigma\right>_{p}$ in the 3D simulation.

For a more quantitative comparison,
Figure \ref{fig:Sigma_min_r1_S8} shows $\left<\Sigma\right>_{\rm gap}$, $\left<\Sigma\right>_{p}$
and $\left<\Sigma\right>_{S8}$ at $r=r_{p}$, versus time. 
Remarkably, the amplitude of the temporal variations in these three quantities  
is considerably smaller in the 3D case compared to the 2D case.
Beyond $500$ orbits,  
$\left<\Sigma\right>_{\rm gap}$ is,
on average, a factor of $\sim 1.3$ smaller in the 2D case compared to the 3D simulation. 
This result is reasonable to some extent, as the same value of $\mathcal{E}$ is used in
both simulations and, therefore, the torque in the 2D simulation is greater at distances
$\lesssim H$ from the perturber compared to the 3D simulation.
In contrast, the $\left<\Sigma\right>_{p}$ values in the 3D simulation are approximately 
$2.5$ times smaller than in the 2D case. As shown in Figure \ref{fig:Sigma_R_3D},
this difference arises from the radial offset in the gap position and a slight change
in the slope of the radial density profile at the inner edge of the gap. 
Interestingly, while $\left<\Sigma\right>_{p}$ is a factor of $2.5$ smaller
in the 3D case compared to the 2D case, the values of $\left<\Sigma\right>_{S8}$
at $r=r_{p}$, averaged from $200$ to $1100$ orbits, are quite similar.

\section{Summary and conclusions}
\label{sec:conclusions}
 
The gravitational interaction between a satellite on a fixed circular orbit and
a retrograde rotating disc differs fundamentally from the prograde
case. In the prograde scenario, the interaction is governed by Lindblad and 
corotation resonances, whereas no resonances are present in the retrograde
case. The formation and structure of gaps also differs between retrograde
and prograde orbiters. To quantify the depth and eccentricity
of these gaps, examine the relative position of the perturber with respect 
to the gap's minimum, estimate the disc eccentricity and evaluate 
the influence of the IBCs, we have conducted 
a number of 2D simulations. These simulations involve a non-accreting perturber
with $q=0.005-0.015$ on a retrograde orbit within the midplane of a viscous
gas disc ($\nu_{-5}$ was varied between $0.25$ and $4$). 

We find that a viscous criterion for gap formation, derived for perturbers
in circular orbits (Equation \ref{eq:cond_gap_formation}), aligns well with simulation results.
We have also examined the scaling of gap depth, 
$\Sigma_{\rm gap}/\Sigma_{{\rm un},p}$ with $q/q_{\nu}$ 
and identified a trend between them, albeit with some scatter. 

In contrast to the prograde case, the body does not sit at the gap's 
minimum but is located on the inner rim of the gap. As a result, 
$\left<\Sigma\right>_{p}$ can be up to $20$ times larger than 
$\left<\Sigma\right>_{\rm gap}$.

The perturber, even when fixed on a circular orbit, drives radial motions
in the disc, causing temporal variations in both 
$\left<\Sigma\right>_{\rm gap}$ and $\left<\Sigma\right>_{p}$ over an orbital period. 
In addition, the perturber can excite eccentric modes in the disc.
As a consequence, the surface density ahead of the body, $\left<\Sigma\right>_{S8}$,
may be highly fluctuating. However, we have found that the level of
disc eccentricity excitation, and thus the amplitude of the
fluctuations in $\left<\Sigma\right>_{S8}$ is sensitive to the IBCs. We have noticed that,
over sufficiently long times, the disc eccentricity may be dominated
by spurious effects generated at the inner boundary. Special
care should be taken with simulations where $q>0.015$ or $\mathcal{E}\leq 0.3$,
as these boundary effects might lead to unrealistic results after a 
few hundred orbits.

We find that in the freely migrating case, the migration is faster in discs with higher 
viscosity. The gravitational coupling between the satellite and the disc lead to the excitation of the satellite's orbital eccentricity up to values comparable to the disc 
aspect ratio $h$.

We have also conducted a 3D simulation with $q=0.01$ and $\nu_{-5}=1$ to investigate
the meridional flow in the disc. Like in the 2D simulations, the perturber is
located at the inner edge of the gap. Behind the perturber (in sector S1),  the
gravitational pull of the perturber compresses the gas flow radially and vertically. 
This compression accelerates radially along the gap and collides
supersonically with the inner edge of the gap. Behind this compression tail (sectors
S3 to S7), the gas at the gap's edge moves upward away from the midplane. 

We have compared $\left<\Sigma\right>_{\rm gap}$, $\left<\Sigma\right>_{p}$
and $\left<\Sigma\right>_{S8}$ from the 3D simulation to those from the 2D
simulation. Our results show that $\left<\Sigma\right>_{\rm gap}$ is slightly higher
in the 3D simulation, while $\left<\Sigma\right>_{p}$ is $\sim 2.5$ times
lower in the 3D simulation. Additionally, the amplitude
of the temporal varations in $\left<\Sigma\right>_{S8}$ is reduced in the 3D 
compared to the 2D simulation.

More realistic simulations should include the accretion of mass onto the perturber,
as in \citet{iva15}. Another limitation of our 3D simulation is that we imposed reflection
symmetry between the upper and lower halves of the disc, which may suppress
the onset of vertical instabilities. We plan to perform fully 3D 
simulations to explore whether the tilting instability found by \citet{ove24} can
also arise in the case of embedded perturbers. 

\section*{Acknowledgements}
We thank the referee for thoughtful comments and suggestions.
The numerical calculations were performed on the Miztli supercomputer at UNAM (project LANCAD-UNAM-DGTIC-03).

\section*{Data Availability}
The FARGO3D code is available from  https://github.com/FARGO3D/fargo3d.
The output files of our hydrodynamical simulations will be shared on reasonable request to the corresponding author.

\appendix
\section{Torque on the perturber in the linear case}
\label{app:linear_formula}
Two-dimensional disc models, which ignore the vertical size of the disc, are useful to quantify
the contribution to the torque of density perturbations at distances $\gtrsim H$ from the
perturber, which we denote
by $\Gamma_{\rm 2D}$.  It is convenient to introduce $F_{{\rm DF}, 0}$, defined as 
the dynamical friction force on a perturber moving supersonically in rectilinear orbit in a 
2D slab of constant density $\Sigma_{0}$. Using linear theory, \citet{mut11} find that 
\begin{equation}
\vecF_{{\rm DF},0}= - \frac{\pi \Sigma_{0} (G M_{p})^{2}}{R_{\rm soft} V_{\rm rel}^{2}}
\left(\frac{\vecV_{\rm rel}}{V_{\rm rel}}\right),
\end{equation}
where $\vecV_{\rm rel}$ is the relative velocity of the perturber with respect to the gas.
As said before, for a retrograde perturber in circular orbit, $V_{\rm rel}\simeq
2\Omega_{p}r_{p}$ and, therefore, 
\begin{equation}
F_{\rm DF,0}=\frac{\pi}{4} \frac{q^{2} \Sigma_{p} \Omega_{p}^{2} r_{p}^{4}}{R_{\rm soft}}
\end{equation}
in this case. $\Sigma_{p}$ represents the disc surface density at the orbital radius of
the perturber. It is useful to discuss the results in terms
of the reference torque $\Gamma_{0}$, defined as $\Gamma_{0}=-r_{p}  F_{{\rm DF},0}$.

\citet{san18} show that, for
low-mass perturbers, $\Gamma_{\rm 2D}$ presents large oscillations with time
because the perturber catch its own wake. 
In fact, during the time that the perturber takes to reach its own wake ($\pi/\Omega_{p}$),
the sound can only travel a distance $\pi H$, implying that the wake has not had time
to spread. As a consequence of the interaction with
its own wake, the time-averaged torque $\bar{\Gamma}_{\rm 2D}$
is smaller than $\Gamma_{0}$ \citep[see Figure 6 in][]{san18}.
The formula for the torque derived in \citet{iva15}
in linear theory predicts correctly the time averaged torque, and it can be approximated by
\begin{equation}
\bar{\Gamma}_{\rm 2D} = \lambda _{\mathcal{E}}\Gamma_{0},
\label{eq:lambda_Gamma}
\end{equation}
where
\begin{equation}
\begin{aligned}
\lambda_{\mathcal{E}} & \simeq  \frac{8\mathcal{E} }{\sqrt{3}}
\biggl(\exp\left[-2\sqrt{3}\mathcal{E}\right]+\frac{2}{\sqrt{5}}
\exp\left[-2\sqrt{5}\mathcal{E}\right] \\
&+\frac{\sqrt{3}}{2} \frac{\exp(-12\mathcal{E})}{1-\exp(-4\mathcal{E})}\biggl),
\label{eq:ivanov_approx}
\end{aligned}
\end{equation}
where $\mathcal{E}=R_{\rm soft}/H$ \citep{san18}.
For a disc aspect ratio $h=0.05$ and a softening parameter $\mathcal{E}=0.6$, 
$\lambda_{\mathcal{E}}=0.5$.

\section{Migrating perturbers}
\label{sec:migrating}
The backreaction of an eccentric disc on the
perturber may lead to the excitation of the perturber's eccentricity.
In this subsection we present the evolution of the semi-major axis $a_{p}$ and orbital
eccentricity $e_{p}$ of the perturber when it is allowed to migrate freely under the 
influence of the gas drag.  Initially, the perturbers are on a circular 
orbit with radius $r=1$. The mass of the disc interior to this radius is 
$2.6\times 10^{-3}M_{\bullet}$. We first verified that, for $q=0.001$ (linear case),
the orbital migration rate matches the value predicted using the torque specified in Eq.
(\ref{eq:lambda_Gamma}).

\begin{figure}
\hskip -0.1cm
\includegraphics[angle=0,width=92mm,height=69mm]{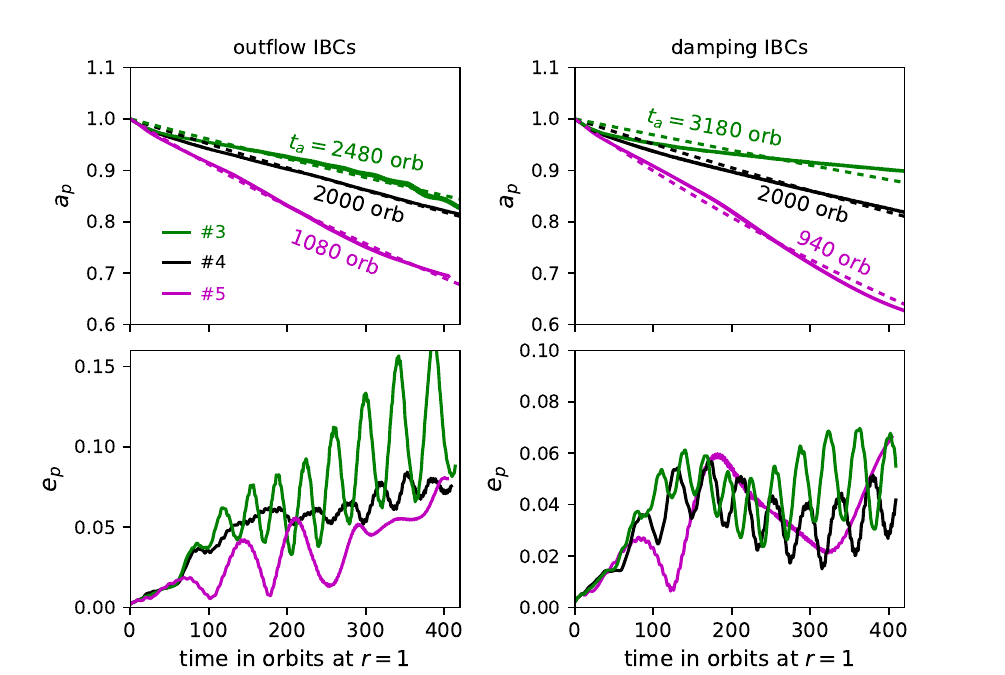}
\vskip -0.05cm
 \caption{Semi-major axis (top panels) and eccentricity of the perturber (bottom panels)
for migrating perturbers as obtained in the 2D simulations (solid lines). The dashed lines
in the top panels
represent an exponential decay fit: $a_{p}=\exp(-t/t_{a})$.  The values of $t_{a}$ are
provided as labels for each curve.
  }
 \label{fig:migrating}
\end{figure}

Figure \ref{fig:migrating}
shows the evolution of $a_{p}$ and $e_{p}$ in models 3, 4 and 5, using 
outflow and damping IBCs. We have fitted the temporal evolution of
the semi-major axis with an exponential function $a_{p}(t)=
\exp (-t/t_{a})$ (see dashed curves in the top panels of Fig. \ref{fig:migrating}).
Regardless of whether the IBCs are outflow or damping, the migration rate, $t_{a}^{-1}$,
is higher in more viscous discs, as expected \citep[e.g.,][]{iva15}. Unlike in model 4, 
where the perturber migrates at the same rate with both outflow and damping IBCs, in 
models 3 and 5, the migration rate depends on the adopted IBCs.

It is remarkable that non-axisymmetric modes in the disc can excite the eccentricity
of the perturber to values comparable to $h$. In models 3 and 4,
the orbital eccentricity increases up to $\sim 0.04$ within the first $100$ orbits, in both
outflow and damping IBCs. The subsequent evolution of the orbiter's
eccentricity, however, is sensitive to the choice of IBCs. 
The variations in $e_{p}$ are caused by the phasing
of the orbital eccentricity with respect to the disc eccentricity.

\bsp	\label{lastpage}
\end{document}